\documentclass[aps,prc,reprint,groupedaddress]{revtex4-1}
\usepackage{float}
\usepackage{graphicx}
\usepackage{epsfig}
\usepackage{geometry}
\usepackage{fancyhdr,amsmath,amssymb,hyperref}
\usepackage{longtable}
\usepackage{hhline}
\usepackage{cleveref}
\usepackage{amsmath}
\usepackage{verbatim}
\begin{document}

%\preprint{}
\title{The fast neutron induced fission of $^{240}$Pu and $^{242}$Pu}



\author{A. Pica$^1$}
\author{A.T. Chemey$^{2}$}
 \author{W. Loveland$^1$}
\affiliation{ 1. Department of Chemistry, Oregon State University, Corvallis, Oregon 97331 USA}
\affiliation{2. Department of Nuclear Science and Engineering,Oregon State University,Corvallis, Oregon 97331 USA}

\date{\today}

\begin{abstract}
We report the measurement of the TKE release in the fast neutron induced fission of $^{240}$Pu and $^{242}$Pu. The results are compared to the predictions of the GEF model, the CGMF model, and the model of Denisov and Sedykh as well as previous experimental work on these reactions. Our absolute measurements of the TKE release are in good agreement with the previous measurements of Nethaway \textit{et al.} for the interaction of 14.8 MeV neutrons with $^{240}$Pu \cite{Nethaway} and of Winkelmann and Aumann for the interaction of 15 MeV neutrons with $^{242}$Pu \cite{Winkelmann}. The general trends of the measured TKE values agree with phenomenological models but the variances of the TKE distributions are significantly less than predicted by various models. The mean post neutron emission TKE release decreases non- linearly with increasing neutron energy and can be represented as TKE(MeV) = 175.8 $\pm$ 0.3 - (2. 4 $\pm$ 0.8)log$_{10}$E$_n$ - (1.4 $\pm$0.4)log$_{10}$E$_{n}^2$ for $^{240}$Pu and TKE(MeV) = 177.1 $\pm$ 0.3 - (1.2 $\pm$ 0.9)log$_{10}$E$_n$ - (1.8 $\pm$0.5)log$_{10}$E$_{n}^2$ for $^{242}$Pu.

\end{abstract}

\maketitle

\section{Introduction}

Nearly 80 $\%$ of the total prompt energy release in fission occurs in the form of the total kinetic energy (TKE) of the fragments \cite{Unik}. Fission is a large scale collective motion of a few hundred nucleons and as such, represents a challenge to understand. The magnitude of the TKE release in fission depends on the Coulomb repulsion between the nascent fragments at scission and the conversion of the fragment motion at scission into kinetic energy.  TKE is thus an important fundamental feature of the fission process that helps to describe the large scale collective motion of the nucleus. There are practical applications of these processes as the kinetic energies of the fragments are important in the design of nuclear weapons and fast-spectrum nuclear reactors.

\subsection{Previous work on the TKE release in fission}

The TKE release has been measured for various cases of spontaneous fission, thermal neutron induced fission, and energetic fission \cite{Younes}. General information about the TKE release in the neutron induced fission of $ ^{235} $U, $^{238}  $U and $^{239} $Pu for E$_n$ $<$ 20 MeV is discussed in the work of Madland \cite{Madland} and Lestone and Strother \cite{Lestone}.  Additionally, TKE, fragment mass distributions, and other neutron induced fission observables for $ ^{232} $Th \cite{King,Maslov2} , $^{233}  $U \cite{Higgins,Liu} , $^{238}  $U \cite{Duke,Maslov}, $^{237}  $Np \cite{Pica}, $^{239}  $Pu \cite{Meierbachtol, Chemey,Liu,Maslov} have been reported.  The number and scope of studies detailing the induced fission of $^{240} $Pu or $^{242} $Pu remains extremely limited; these studies are summarized in Table \ref{tab:Prev}.

\begin{table}[H]
\footnotesize
\caption{Previous studies on the induced fission of  $^{240,242} $Pu }
\centering
\begin{tabular}{ccc}
\hline\hline
Reaction &E$_n$[MeV] & Reference\\
\hline 
$^{240} $Pu(n,f)     & 1.3   				& \cite{Vorobeva}\\
$^{240} $Pu(n,f) 	& 14.8 				& \cite{Nethaway}\\
$^{242} $Pu(n,f) 	& 1.10 				& \cite{Vorobeva}\\
$^{242} $Pu(n,f) 	& 15.1 				& \cite{Winkelmann}\\
$^{242} $Pu(p,f) 	& 13, 20, 55 		&\cite{Rubchenya}\\
\hline
\end{tabular}
\label{tab:Prev}
\end{table}

In Section II of this paper, we discuss the experimental methods used while in Section III we report the results of the measurements and discuss the implications of our data in comparison with current theoretical models of fission.  We present our conclusions in Section IV.

\section{Experimental Details}
\subsection{Overview of Experiment}

This experiment was carried out using the 15R beam line at the Los Alamos National Laboratory, Los Alamos Neutron Science Center at the Weapons Neutron Research (LANSCE-WNR) facility over a 19 day period in November-December 2021. The experimental arrangements were the same as that used in our previous studies \cite{Pica, Chemey} of the fast neutron induced fission of $^{237} $Np and $^{239} $Pu . “White spectrum” neutron beams were generated from an unmoderated tungsten spallation source using the 800 MeV proton beam from the LANSCE linear accelerator. The experiment was located on the 15R beam line (15$ ^{o} $-right with respect to the proton beam). The fast neutron beam intensities were $\sim$ 10$  ^{5}$ - 10$  ^{6}$/s for E$  _{n}$ = 2 - 100 MeV. The proton beam is pulsed allowing one to measure the time of flight (energy) of the neutrons arriving at the experimental area. The proton beam consists of a 625 $ \mu $s macropulse containing about 340 micropulses of width 250 ps that are spaced 1.8 $\mu$s apart. The macropulses had a repetition rate of 100 Hz. With neutron beam intensities $\sim$ 10$  ^{5}$-10$  ^{6}$ n/s, one must use large solid angles for the detectors or long observation times or both to obtain statistically meaningful data. The total times for the irradiation of $^{240}  $Pu and $^{242}$Pu were 6.5 and 12.8 days, respectively.

\subsection{Neutron beam/general setup}
\begin{figure}[hbp]
\includegraphics[trim = {0 10cm 0 10cm},clip,width = 1\columnwidth]{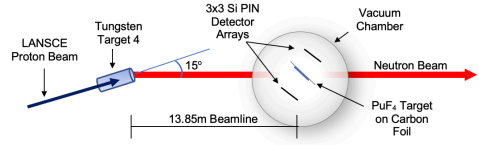}
%trim = {left lower right upper}
\caption{A rough sketch of the experimental setup. (Color online)}
\label{fig:Apparatus}
\end{figure}

The spallation neutrons from the LANSCE tungsten target traversed a 13.85 m flight path to the target position of our scattering chamber. The neutron beam was collimated to a 1.0 cm diameter at the entrance to the experimental area. A fission ionization chamber \cite{Wender} was used to continuously monitor the absolute neutron beam intensities. The targets and the fission detectors were housed in an evacuated, thin- walled aluminum scattering chamber. The scattering chamber was located 84 cm from the collimator, and $\sim$ 14 m from the neutron beam dump. A rough sketch of the experimental setup is shown in Figure \ref{fig:Apparatus}.

\subsection{Targets}

The $^{240}$Pu and $^{242}$Pu targets were prepared by vapor deposition using the method described in \cite{Silveira}. The targets consisted of 2 cm diameter circular deposits of the Pu isotope on a thin C backing. The target thicknesses were 72.3 $\mu$g/cm$  ^{2}$ ($^{240}$Pu) and 63.0 $\mu$g/cm$  ^{2}$ ($^{242}$Pu) (as measured by alpha spectroscopy).  The source material for the $^{240}$Pu target was 99.874 weight percent $^{240}$Pu and the $^{242}$Pu target was 99.964 weight percent $^{242}$Pu. The assumed chemical composition of the targets was that of PuF$_4$. The C backing thickness was 100 $\mu$g/cm$^2$.  The targets were at an angle of 45$ ^{o} $ with respect to the incident beam.

\subsection{Fission detectors}

Fission fragments were detected by 9 pairs of 1 cm$^2$ Si PIN diode detectors (Hamamatsu S3590-09) positioned 3.5 cm from the target as shown in Figure 2.  The time of flight of each interacting neutron was measured using a timing pulse from a Si PIN diode and the accelerator RF signal.  The distance from the spallation target to the Pu targets was measured to be 1384.9 $\pm$ 1.1 cm.  The position of the photo-fission peak in the time of flight spectrum was measured to have an uncertainty less than  2`$\% $.  The neutron energies were calculated using relativistic relationships from the distance traveled by the neutrons, the mid-point of the photo-fission peak in the fission time of flight spectrum and the observed time difference between the neutron timing signal and the accelerator RF signal.  The uncertainties from each of the components of the neutron energy were added in quadrature as uncorrelated uncertainties to determine the final uncertainty in the measured neutron energy. The neutron energies were thus determined with an uncertainty of $\leq$ 2 $\%$. The neutron energies were generally binned logarithmically to give bins of equal associated uncertainty  in the neutron energy.  The width of these bins give the neutron energy resolution of our measurements.

\begin{figure}[hbp]
\centering
\includegraphics[trim = {1cm 5cm 1cm 8cm, },clip,width=1 \columnwidth ]{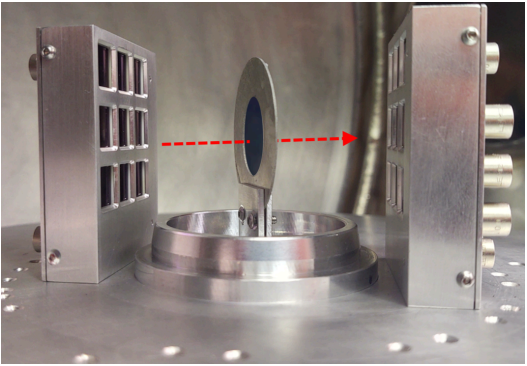}
%trim = {left lower right upper}
\caption{A photo of the fission fragment/target geometry used in the experiment. (Color online)}
\end{figure}

\subsection{The 2E Method}
Determination of TKE and fission product masses must include considerations for the many complex phenomena occurring during the fission process such as multichance fission,  pre-scission neutron emission,  and prompt neutron evaporation.  To account for these phenomena we use the  2E method to calculate fission product  masses based on their kinetic energies within a momentum conserving framework.  (We use the convention that the term "fission fragments” refers to pre-neutron emission while the term “fission product" refers to post neutron emission \cite{Madland}.)

Due to the subfemtosecond timescale of neutron evaporation and the nanosecond flight times of fragments hitting detectors,  we detect the kinetic energies of the post-evaporation products.  An iterative procedure is employed to calculate pre-neutron evaporation fragment masses on an event-by-event basis.   Conservation of momentum and nucleon number dictate: 

\begin{equation}
A^*_HE^*_H = A^*_LE^*_L
\label{eq:1}
\end{equation} 

\begin{equation}
A^*_L + A^*_H = A_{CN}
\label{eq:2}
\end{equation} 

\begin{equation}
A_L^* = \frac{E_H^*} {E_{H}^*+E_L^*}     A_{CN}
\label{eq:8}
\end{equation} 

\noindent where $H$ represents the heavy fragment, $L$ represents the light fragment, CN represents the compound nucleus, and “ $*$ “ indicates a quantity that pertains to pre-neutron emission.  A$_i^*$ is the mass of fragment \textit{i}. A$_i^*$ (E$_i^*$) and post neutron A$_i$ (E$_i$) are related through

\begin{equation}
A_i = A_i^* - {v_{post} (A_i^*,E_n)}
\label{eq:3}
\end{equation}

\noindent where ${v_{post} (A)}$ is the post fission prompt neutron multiplicity calculated from the General Description of Fission Observables (GEF) model \cite{GEF1, GEF2} or the Cascade Gamma Multiplicities from Fission (CGMF) model \cite{CGMF}.  Assuming isotropic neutron emission (which leaves the average fragment velocities unaffected by neutron emission),  the post-neutron energy is calculated by:

\begin{equation}
E_{i} = \frac{A_{i}} {A_{i}^*}     E_{i}^*
\label{eq:4}
\end{equation}

\noindent which,  re-written using equation (\ref{eq:3}) is

\begin{equation}
E^*_{i} = E_{i} ( 1 + \frac{v_{post}(A^*_{i})}{A_{i}})
\label{eq:7}
\end{equation}

\noindent Combining equations (1-5)  we get the following relation between pre- and post-neutron masses:

\begin{equation}
A_H^* \approx \frac{A_{CN}E_L}{E_H+E_L(1+\xi_H)}
\label{eq:5}
\end{equation}

\begin{equation}
\xi_H \equiv  \frac{1+v_H/A_H}{1+v_L/A_L} - 1 
\label{eq:6}
\end{equation}

\noindent If we assume no neutron emission,  the provisional fragment mass $\mu_i$,  can be calculated by simplifying equation (\ref{eq:5}) to:

\begin{equation}
\mu_H = A_{CN} \frac{E_L}{E_H+E_L}
\end{equation}

\noindent Once $\mu_i$ are calculated, they are compared to the pre-neutron evaporation fragment masses of the previous iteration.  Corrections for PHD, the pulse-height defect, and energy loss of fragments in the target and backing foils are also applied to this iterative procedure.  This continues until $\mu_H$ and $\mu_L$ have converged to the pre-neutron fragment masses, defined here as an iterative mass difference of less than 0.1 u. 

As previously mentioned, the GEF and CGMF models were used to quantify the impact of multichance fission.  Multichance fission occurs when the excitation energy of the compound nucleus is greater than the neutron separation energy. If this condition is met, the compound nucleus can emit pre-scission neutrons which lowers the available energy to transfer to the fragments at scission. Due to the phenomenon of multichance fission an ensemble of nuclides contributes to the overall fission observables.  The contributions of each CN as predicted by GEF in the multichance fission chain for the $^{240, 242}$Pu (n,f) reaction are given in Tables \ref{tab:MCF1} and \ref{tab:MCF2} \cite{GEF1}.

\begin{table*}
\footnotesize
\caption{Contribution of each compound nucleus for a given E$_n$ for $^{240}$Pu(n,f), calculated from GEF \cite{GEF1}  }
\centering
\begin{tabular}{cccccccccccc}
\hline\hline
E$_n$[MeV] &241$(\%)$ & 240$(\%)$ & 239$(\%)$ & 238$(\%)$ & 237$(\%)$ & 236$(\%)$ & 235$(\%)$& 234$(\%)$\\
\hline 
0.8	&100	&-&-	&-	&-&	-&	-&	-\\
1.2	&100	&-&	-	&-	&-&	-	&-	&-\\
1.8	&100	&-	&-	&-	&-	&-	&-	&-\\
2.6	&100	&-	&-	&-	&-	&-	&-	&-\\
3.8	&100	&-	&-	&-	&-	&-	&-	&-\\
6.0	&78		&22	&-	&-	&-	&-	&-	&-\\
9.5	&53		&47	&-	&-&	-&	-	&-	&-\\
17.9	&21		&44&	33	&-&	-	&-	&-&	-\\
36.3	&7	&21	&26&	30	&16	&-&	-&	-\\
73.4			&1&6&	9	&14&	17&	21	&21&	11\\
\hline
\end{tabular}
\label{tab:MCF1}
\end{table*}

\begin{table*}
\footnotesize
\caption{Contribution of each compound nucleus for a given E$_n$ for $^{242}$Pu(n,f), calculated from GEF \cite{GEF1}  }
\centering
\begin{tabular}{ccccccccccccc}
\hline\hline
E$_n$[MeV] &243$(\%)$ & 242$(\%)$ & 241$(\%)$ & 240$(\%)$ & 239$(\%)$ & 238$(\%)$ & 237$(\%)$& 236$(\%)$&235$(\%)$\\
\hline 
0.9	&100	&-	&-	&-	&-	&-	&-	&-	&-\\
1.3	&100	&-	&-	&-	&-	&-	&-	&-	&-\\
1.9	&100	&-	&-	&-	&-	&-	&-	&-	&-\\
2.8	&100	&-	&-	&-	&-	&-	&-	&-	&-\\
4.2	&100	&-	&-	&-	&-	&-	&-	&-	&-\\
6.7	&49	&51	&-	&-	&-	&-	&-	&-	&-\\
11.2	&43	&57	&-	&-	&-	&-	&-	&-	&-\\
21.9	&15	&30	&38	&17	&-	&-	&-	&-	&-\\
42.5	&4	&12	&18	&24	&27	&25	&-	&-	&-\\
76.5	&1	&3	&6	&10	&14	&18	&20	&21	&7\\
\hline
\end{tabular}
\label{tab:MCF2}
\end{table*}

\subsection{PHD and Fragment Energy Loss}
We measured, on an event-by-event basis,  the pulse heights of coincident fission fragments to determine their energies.  However,  incomplete charge collection, recombination, and non-ionizing collisions lead to unavoidable pulse height defects.  Individual detectors were calibrated by the Schmitt method \cite{Schmitt} with a spontaneously-fissioning $^{252}$Cf source, using the updated Weissenberger calibration parameters \cite{Weissenberger}. Pulse height defect varies with energy and mass of fission fragment, and must furthermore be inferred for each individual detector. For Si surface barrier detectors, the magnitude of the PHD generally ranges from 3-13 MeV \cite{Ogihara,Kitahara,Kaufman}.  PHD is mass- and energy-dependent, and is only implicitly known from detector energy calibration by the Schmitt method. The approximate PHD for our PIN diode array was 4.8-7.7 MeV, as obtained from the constant-offset term of the energy calibration \cite{Schmitt}. This is consistent with analyses that resolve both energy and mass of implantations \cite{Knoll}. The energy loss of the fission products in the target deposit and the carbon foil were calculated using the Northcliffe-Schilling tables and assuming that the fission event took place in the middle of target \cite{Northcliffe}. The linear momentum transfer (LMT) was parameterized using data from the $^{239}$Pu(n,f) reaction and fit from E$_n$ = 7-120 MeV \cite{Hensle1,Hensle2}.  For E$_n$ $<$ 7 MeV,  LMT was assumed to be 100 $\%$. 

\subsection{Neutron Multiplicity}
The compound nucleus is cooled considerably by the emission of pre-equilibrium and pre-scission neutrons. Our experimental  apparatus measures the kinetic energies of coincident fission products.  To deduce the mass and energy of the fission fragments we must account for prompt neutron emission. To a good approximation, the total excitation energy of a given fission fragment is the sum of the total available intrinsic excitation energy at scission, and the distortion energy of the fragment at scission. The sum of these two contributions determines the energy available for neutron evaporation.  Experimental data outlining the behavior of prompt neutron emission, $v(A)$, is extremely scarce for many isotopes at any $E_n$.  Accordingly,  most studies of this nature utilize Monte-Carlo simulations of  fission fragment de-excitation after scission to determine average neutron multiplicity. 

\begin{figure}
 \includegraphics[trim={0 0 0 0 },clip,width=1\columnwidth]{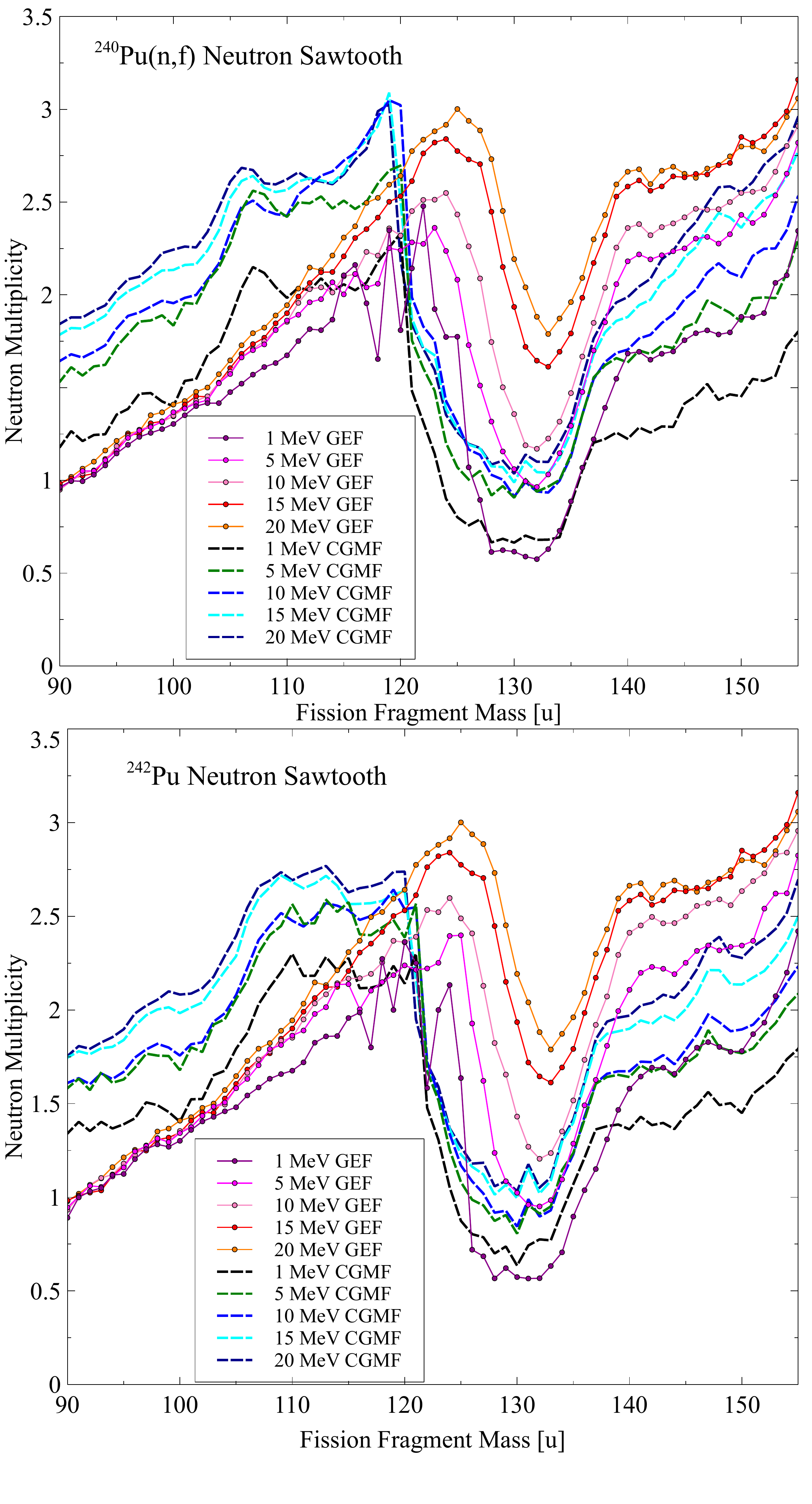}
\caption{Average neutron multiplicity for $^{240}$Pu (top) and $^{242}$Pu (bottom) at several incident neutron energies calculated using GEF \cite{GEF1} and CGMF \cite{CGMF}.  (Color online)}
\label{fig:Sawtooth}       % Give a unique label
\end{figure}

For the present analysis,  neutron multiplicities for E$_n$ = 1 - 20 MeV from two fission models, GEF \cite{GEF1} and CGMF \cite{CGMF}, were used to find the pre and post neutron evaporation masses from the measured fission product energies.  For E$_n > $ 20 MeV,  only the GEF neutron multiplicities were used in the 2E analysis as neutron induced fission of E$_n >$ 20 MeV is not yet modeled by CGMF.  The neutron multiplicities for $^{240}$ Pu(n,f) and $^{242}$ Pu(n,f) obtained by GEF and CGMF are shown in Figure \ref{fig:Sawtooth}.  It is apparent that the GEF and CGMF codes divide the energy of the incident neutron quite differently.

Several theories exist \cite{Schmidt3},\cite{Becker},\cite{Litaize},\cite{Morariu} to describe the partitioning of energy between two fission fragments. The physics behind the GEF model v(A) values is described in \cite{Schmidt3}.  Within this framework the initial excitation of the compound nucleus is shared by the protofragments according to the ratio of their masses. However, when pairing correlations are added to the regime, energy sorting takes place, resulting in a total transfer of excitation energy from the light fragment to the heavy fragment \cite{Schmidt1}\cite{Schmidt2}.  At higher energies, pairing correlations get washed out and there is a transition back to an energy distribution closer to the ratio of the fragment masses. This phenomenon of energy sorting dictates that the additional energy brought by the incident particle is given to the heavy fragment and thus raises the neutron multiplicities of the heavy fragments only \cite{Schmidt3}.

The CGMF model, on the other hand, uses the approach of \cite{Becker} to predict v(A).
Within this framework it is the ratio of the initial fragment temperatures, R$_T$, that determines the partitioning of the excitation energy.  The ratio of nuclear temperatures is correlated to the degree of fragment deformation. The heavy fragments formed near the closed shell of Z = 50 and N = 82 are expected to be more spherical than their lighter counterparts. As such, the light fragment is expected to acquire more of the bombarding particles excitation resulting in higher neutron multiplicities of the light fragment.  Despite the different approaches to modeling v(A) between GEF and CGMF there is virtually no difference between the predicted total neutron multiplicity (Table \ref{tab:neutron}).

\begin{table}
\footnotesize
\caption{Mean total neutron multiplicity [v(A)] predicted by GEF \cite{GEF1} and CGMF \cite{CGMF} for $^{240,242}$Pu (n,f) }
\centering
\begin{tabular}{ccccc}
\hline\hline
E$_n$[MeV] &$^{240}$GEF     & $^{240}$CGMF & $^{242}$GEF& $^{242}$CGMF \\
\hline 
1		&	2.70	&	2.78	& 2.78	& 3.01 	\\
5		&	3.27	&	3.59	& 3.37	& 3.55 \\
10		&	4.07	&	4.29	& 4.17	& 4.21 	\\
15		&	4.81	&	4.95	& 4.90	& 4.86	\\
20	&	5.3	&	5.52	& 5.44	& 5.57	\\
\hline
\end{tabular}
\label{tab:neutron}
\end{table}

%\begin{figure}
%% Use the relevant command to insert your figure file.
%% For example, with the graphicx package use
% \includegraphics[width=.75\textwidth]{Sawtooth_240}
%\caption{Average neutron multiplicity for Pu-240 (left) and Pu-242 (right) at several incident neutron energies calculated using GEF \cite{GEF1} and CGMF \cite{CGMF}. }
%\label{fig:Sawtooth_240}       % Give a unique label
%\end{figure}
%
%\begin{figure}
%% Use the relevant command to insert your figure file.
%% For example, with the graphicx package use
% \includegraphics[width=.75\textwidth]{Sawtooth_242}
%\caption{Average neutron multiplicity for $^{242}$ Pu(n,f) at several incident neutron energies calculated using GEF \cite{GEF1} and CGMF \cite{CGMF}. }
%\label{fig:Sawtooth_242}       % Give a unique label
%\end{figure}

 \section{Results}
\begin{figure}[!htb]
 \includegraphics[width=1 \columnwidth]{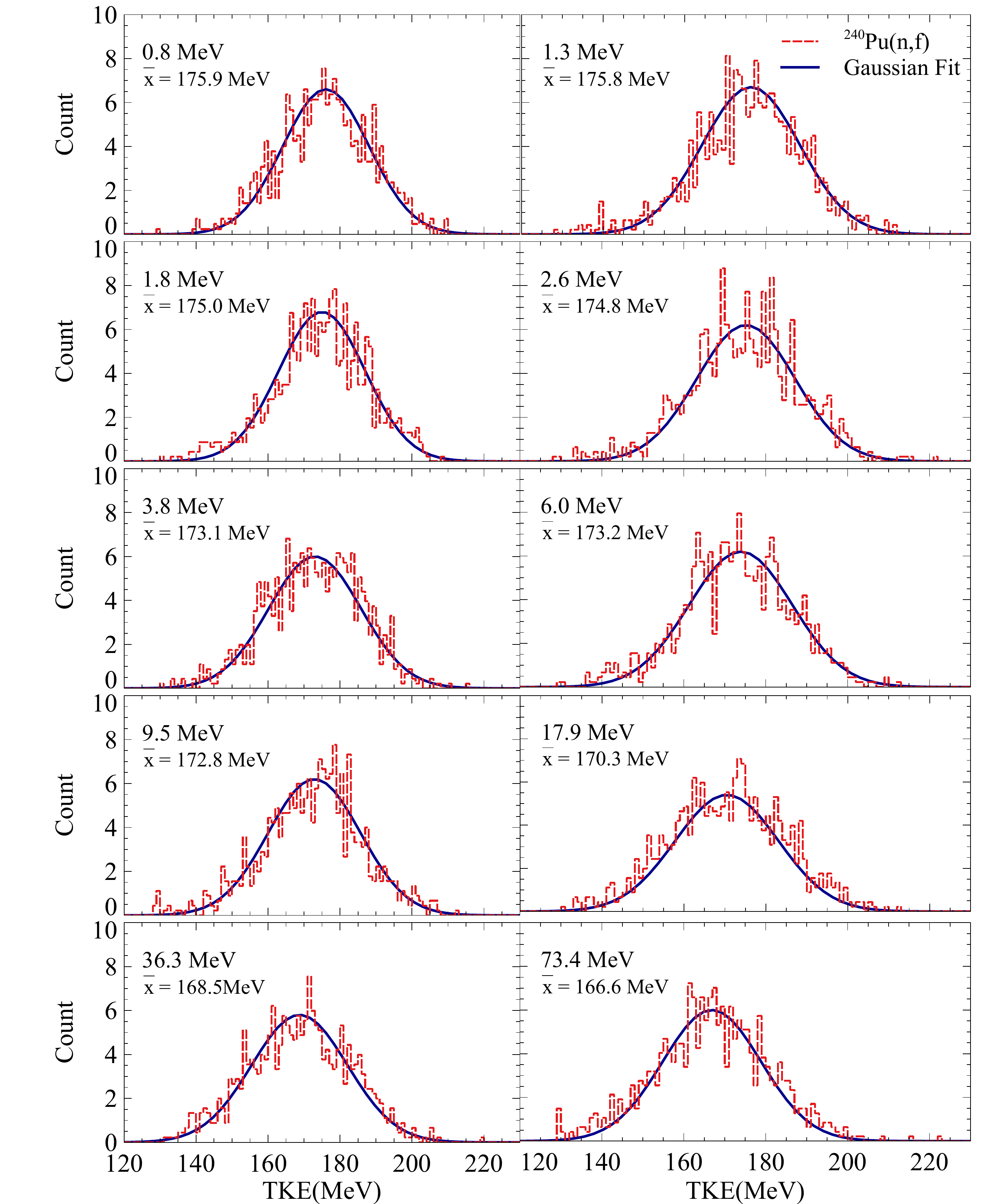}
\caption{Plots for the $^{240}$Pu TKE$_{post}$ distributions sorted into the energy bins described in Table \ref{tab:240TKE}. The plotted curves (solid line) represent the results of fitting the distributions with Gaussian functions. (Color online).}
\label{fig:240TKEvCount}       % Give a unique label
\end{figure}

\begin{figure}[!htb]
 \includegraphics[width=1 \columnwidth]{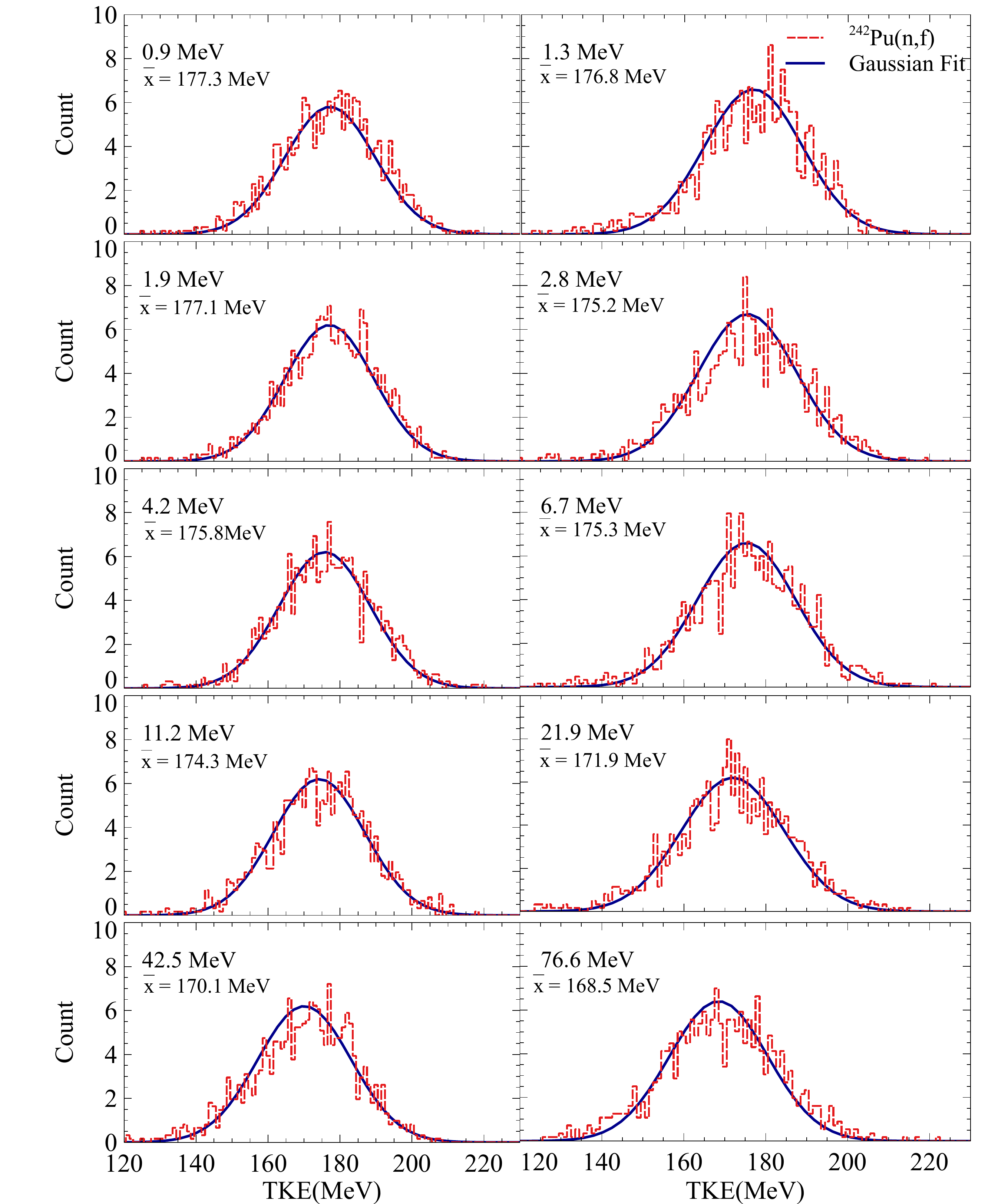}
\caption{Plots for the $^{242}$Pu TKE$_{post}$ distributions sorted into the energy bins described in Table \ref{tab:242TKE}. The plotted curves (solid line) represent the results of fitting the distributions with Gaussian functions. (Color online).}
\label{fig:242TKEvCount}       % Give a unique label
\end{figure}

9,099 coincident fission events for $^{240}$Pu(n,f) and 12,250 coincident fission events for $^{242}$Pu(n,f) were recorded over their one and two week irradiation periods respectively.  In Table \ref{tab:240TKE} and Table \ref{tab:242TKE} we report the pre- and post-neutron TKE and its variance as a function of incident neutron energy for the $^{240}$Pu(n,f) and $^{242}$Pu(n,f) reactions.   In  Figures \ref{fig:240TKEvCount} and \ref{fig:242TKEvCount}, we present the post neutron evaporation TKE distributions sorted by neutron energy bin. The data were binned as to have a similar number of events per neutron energy bin. The solid lines in each plot represent the results of fitting the data (dotted line) with Gaussian distributions.  The variances of these Gaussian distributions ($\sigma^2$) are sensitive indicators of multichance fission, broadening when a new fission channel opens up. 

 \subsection{TKE vs E$_n$}
 
 \begin{table*}
\centering
\caption{TKE and $\sigma^2_{post}$ in the $^{240}$Pu (n,f) reaction as a function of incident neutron energy.  The neutron bin limits are given the first column and the second column is the geometric mean of the neutron energies. The last column is the number of events \textit{N} in the bin.}
\label{tab:240TKE}  
\begin{tabular}{cccccc}
\hline\hline\noalign{\smallskip}
E$_n$ Range(MeV) &  E$_n$ (MeV) & $_{pre}$TKE (MeV) & $_{post}$TKE (MeV) & $\sigma^{2}_{post}$TKE & Events \\
\hline\hline\noalign{\smallskip}\noalign{\smallskip}
\relax[0.55-1.01] & 0.78 & 177.9$\pm$ 0.5 & 175.9$\pm$ 0.5& 161.5$\pm$ 0.4& 847 \\
\relax[1.01-1.47] & 1.25 & 178.0$\pm$ 0.4 & 175.8$\pm$ 0.4& 147.6$\pm$ 0.4& 936 \\
\relax[1.47-2.17] & 1.82 & 177.1$\pm$ 0.5 & 175.0$\pm$ 0.5& 170.6$\pm$ 0.5& 918 \\
\relax[2.17-3.04] & 2.61 & 177.2$\pm$ 0.5 & 174.8$\pm$ 0.5& 157.8$\pm$ 0.4& 933 \\
\relax[3.04-4.64] & 3.84 & 175.1$\pm$ 0.5 & 173.1$\pm$ 0.5& 155.0$\pm$ 0.4& 910 \\
\relax[4.64-7.35] & 6.00 & 175.7$\pm$ 0.5 & 173.2$\pm$ 0.5& 156.3$\pm$ 0.4& 906 \\
\relax[7.35-11.68] & 9.51 & 175.2$\pm$ 0.5 & 172.8$\pm$ 0.5& 172.7$\pm$ 0.5& 901 \\
\relax[11.68-24.21] & 17.94 & 173.2$\pm$ 0.5 & 170.3$\pm$ 0.4& 192.9$\pm$ 0.4& 905 \\
\relax[24.21-48.34] & 36.27 & 171.5$\pm$ 0.5 & 168.5$\pm$ 0.4& 200.2$\pm$ 0.4& 902 \\
\relax[48.34-100] & 73.36 & 170.0$\pm$ 0.5 & 166.6$\pm$ 0.5& 193.7$\pm$ 0.4& 941 \\
\noalign{\smallskip}\hline\hline
\end{tabular}
\end{table*}

\begin{table*}
\centering
\caption{TKE and $\sigma^2_{post}$ in the $^{242}$Pu (n,f) reaction as a function of incident neutron energy.  The neutron bin limits are given the first column and the second column is the geometric mean of the neutron energies. The last column is the number of events \textit{N} in the bin.}
\label{tab:242TKE}  
\begin{tabular}{cccccc}
\hline\hline\noalign{\smallskip}
E$_n$ Range(MeV) & E$_n$ (MeV) & $_{pre}$TKE (MeV) & $_{post}$TKE (MeV) & $\sigma^{2}_{post}$TKE & Events \\
\noalign{\smallskip}\hline\hline\noalign{\smallskip}
\relax[0.62-1.09] & 0.85 & 179.5$\pm$ 0.4 & 177.3$\pm$ 0.4& 174.2$\pm$ 0.3& 1223 \\
\relax[1.09-1.55] & 1.32 & 178.7$\pm$ 0.4 & 176.8$\pm$ 0.4& 149.8$\pm$ 0.3& 1254 \\
\relax[1.55-2.26] & 1.91 & 179.4$\pm$ 0.4 & 177.1$\pm$ 0.4& 157.3$\pm$ 0.3& 1272 \\
\relax[2.26-3.28] & 2.77 & 177.7$\pm$ 0.4 & 175.2$\pm$ 0.4& 188.5$\pm$ 0.3& 1239 \\
\relax[3.38-5.10] & 4.19 & 178.0$\pm$ 0.4 & 175.8$\pm$ 0.4& 164.1$\pm$ 0.3& 1242 \\
\relax[5.10-8.36] & 6.73 & 177.8$\pm$ 0.4 & 175.3$\pm$ 0.4& 157.5$\pm$ 0.3& 1233 \\
\relax[8.36-13.97] & 11.16 & 176.5$\pm$ 0.4 & 174.3$\pm$ 0.4& 144.7$\pm$ 0.3& 1223 \\
\relax[13.97-29.86] & 21.92 & 174.7$\pm$ 0.4 & 171.9$\pm$ 0.4& 184.7$\pm$ 0.3& 1227 \\
\relax[29.86-42.50] & 42.50 & 173.1$\pm$ 0.4 & 170.1$\pm$ 0.4& 185.0$\pm$ 0.3& 1221 \\
\relax[42.50-100] & 76.55 & 171.8$\pm$ 0.4 & 168.5$\pm$ 0.4& 190.4$\pm$ 0.4& 1116 \\
\noalign{\smallskip}\hline\hline
\end{tabular}
\end{table*}

\subsubsection{Comparison with GEF and CGMF}
\begin{figure*}
%\hspace*{-2cm}
 \includegraphics[trim = {0 .5in 0 0in},clip, width=1 \textwidth]{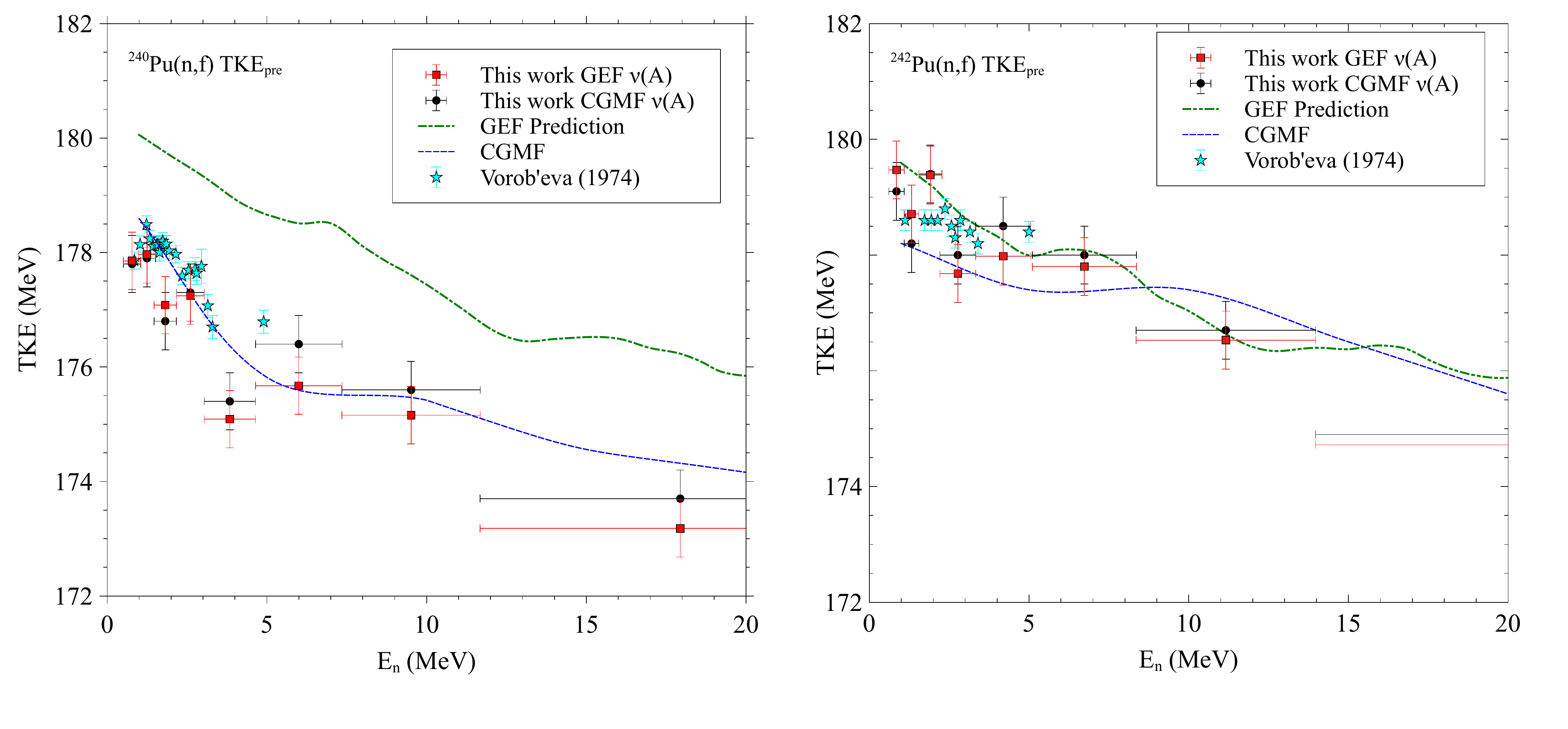}
\caption{ TKE$_{pre}$ v E$_n$ for the $^{240}$Pu(left) and $^{242}$Pu (right) using both the GEF (red squares)\cite{GEF1} and CGMF (black circles)\cite{CGMF} neutron multiplicities. Our data are compared to the GEF TKE$_{pre}$ predictions and CGMF inputs.  The parameterizations in CGMF for $^{240,242}$Pu(n,f) were extrapolated from the well characterized $^{239,241}$Pu(n,f) reactions.  We also show the TKE release measured in Ref \cite{Vorobeva}. (Color online)}
\label{fig:CGMFvGEF}       % Give a unique label
\end{figure*}

Pre-neutron evaporation (TKE$_{pre}$) results for incident neutron energies of E$_n < $ 20 MeV are shown in Figure \ref{fig:CGMFvGEF}. In doing so, we restrict our attention to incident energies where both GEF and CGMF neutron multiplicities were implemented in the analysis regime. In 1974 Vorob'eva \textit{et al.} measured the kinetic energy release for $^{240,242}$Pu(n,f) for E$_{n}<$ 3.5 MeV and reported the mean kinetic energy of fragments for $^{240}$Pu (n,f) and $^{242}$Pu(n,f) to be 178.2 $\pm$ 0.5 MeV and 178.6 $\pm$ 0.5 MeV, at E$_n$ = 1.3 and 1.1 MeV respectively,  relative to the mean kinetic energy release of $^{235}$U by thermal neutrons \cite{Vorobeva}.  These data are presented in Figure \ref{fig:CGMFvGEF}. Notwithstanding the measurements of Vorob'eva \textit{et al.}, data characterizing the $^{240,242}$Pu(n,f) reactions remains extremely scarce.  As such, parameterizations in CGMF for $^{240,242}$Pu(n,f) were extrapolated from the well studied $^{239,241}$Pu(n,f) reactions.  There is no statistical difference between the TKE calculated using GEF v(A) v. CGMF v(A).

\begin{figure*}
 \includegraphics[trim={10mm 0 0 0 },clip,width=1 \textwidth]{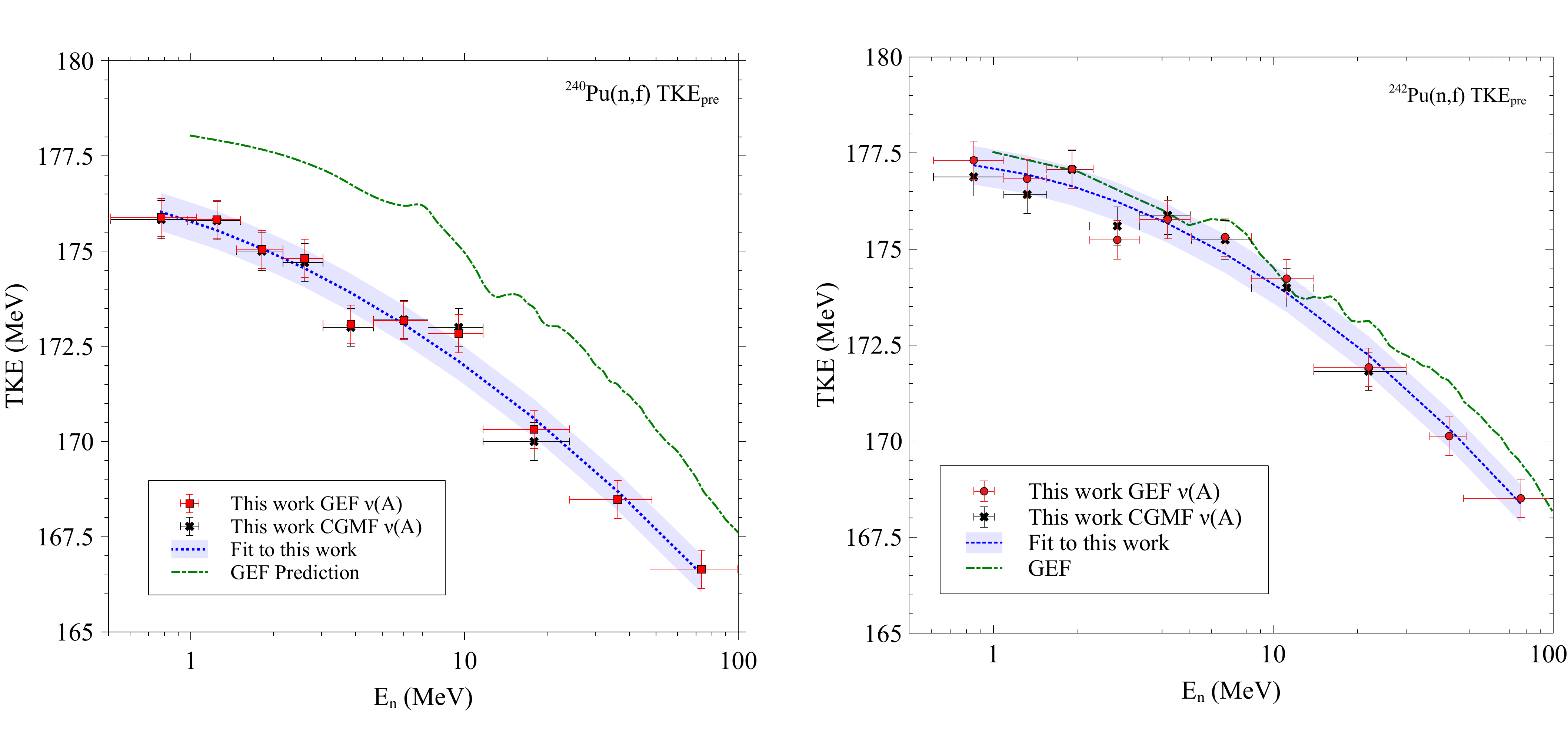}
\caption{TKE$_{post}$ v E$_n$ for the $^{240}$Pu(n,f)  (left) and $^{242}$Pu(n,f) (right) reactions using both the GEF (red squares) \cite{GEF1} and CGMF (black circles) \cite{CGMF} neutron multiplicites from E$_n$ =  1-100 MeV and E$_n$ = 1-20 MeV, respectively. Our data are compared to the GEF TKE$_{post}$ predictions.  TKE error bars are statistical.  The shaded region is the standard error of the fit to this work at a 95$\%$ confidence interval. (Color online) }
\label{fig:PuTKEpost}       % Give a unique label
\end{figure*}

The TKE$_{post}$ release for the fast neutron-induced fission of the Pu isotopes over the entire measured energy range is given in Figure \ref{fig:PuTKEpost}. In this figure we show our data,  the GEF model\cite{GEF1}, and the $^{240,242}$Pu(n,f) data measured by Vorob'eva \textit{et al. }\cite{Vorobeva,Vorobeva2}. The overall trend of decreasing TKE with increasing excitation is consistent with other fast neutron TKE studies of actinides \cite{Pica, Yanez,Higgins,Meierbachtol,Chemey,King}.  A second order log$_{10}$ polynomial fit best describes the TKE vs. E$_n$ relationship. The fit to the $^{240}$Pu data (Figure \ref{fig:PuTKEpost} left) is given by the equation TKE(MeV) = 175.8 $\pm$ 0.3 - (2. 4 $\pm$ 0.8)log$_{10}$E$_n$ - (1.4 $\pm$0.4)log$_{10}$E$_{n}^2$ and the fit to the $^{242}$Pu data (Figure \ref{fig:PuTKEpost} right) is given by the equation TKE(MeV) = 177.1 $\pm$ 0.3 - (1.2 $\pm$ 0.9)log$_{10}$E$_n$ - (1.8 $\pm$0.5)log$_{10}$E$_{n}^2$ at the 95$\%$ confidence interval.  The TKE release for $^{242}$Pu (n,f) is $\sim 1\%$ higher than for $^{240}$Pu (n,f) across the entire energy range.  This stands in contradiction with the general expectation of a linear TKE decrease with increasing A \cite{Viola}, however measurements of spontaneous fission of these two Pu isotopes also find a greater TKE release for $^{242}$Pu (sf)  than for $^{240}$Pu (sf) \cite{Wagemans2,Wagemans3,Dematte}. The discrepancy between the spontaneous and induced fission measurements and the expected Z$^2$/A$^{1/3}$ dependence suggests microscopic effects are predominating over macroscopic behavior \cite{Dematte}. 

The GEF model overestimates the TKE release for $^{240}$Pu (n,f) by $(\sim$ 1.5 MeV) $<1\%$ which is consistent with the accuracy of GEF TKE predictions reported in \cite{Schmidt3} for other, more well characterized, fission systems.

\begin{figure}[!htb]
 \includegraphics[trim={0 1cm 0 0 },clip,width=1\columnwidth]{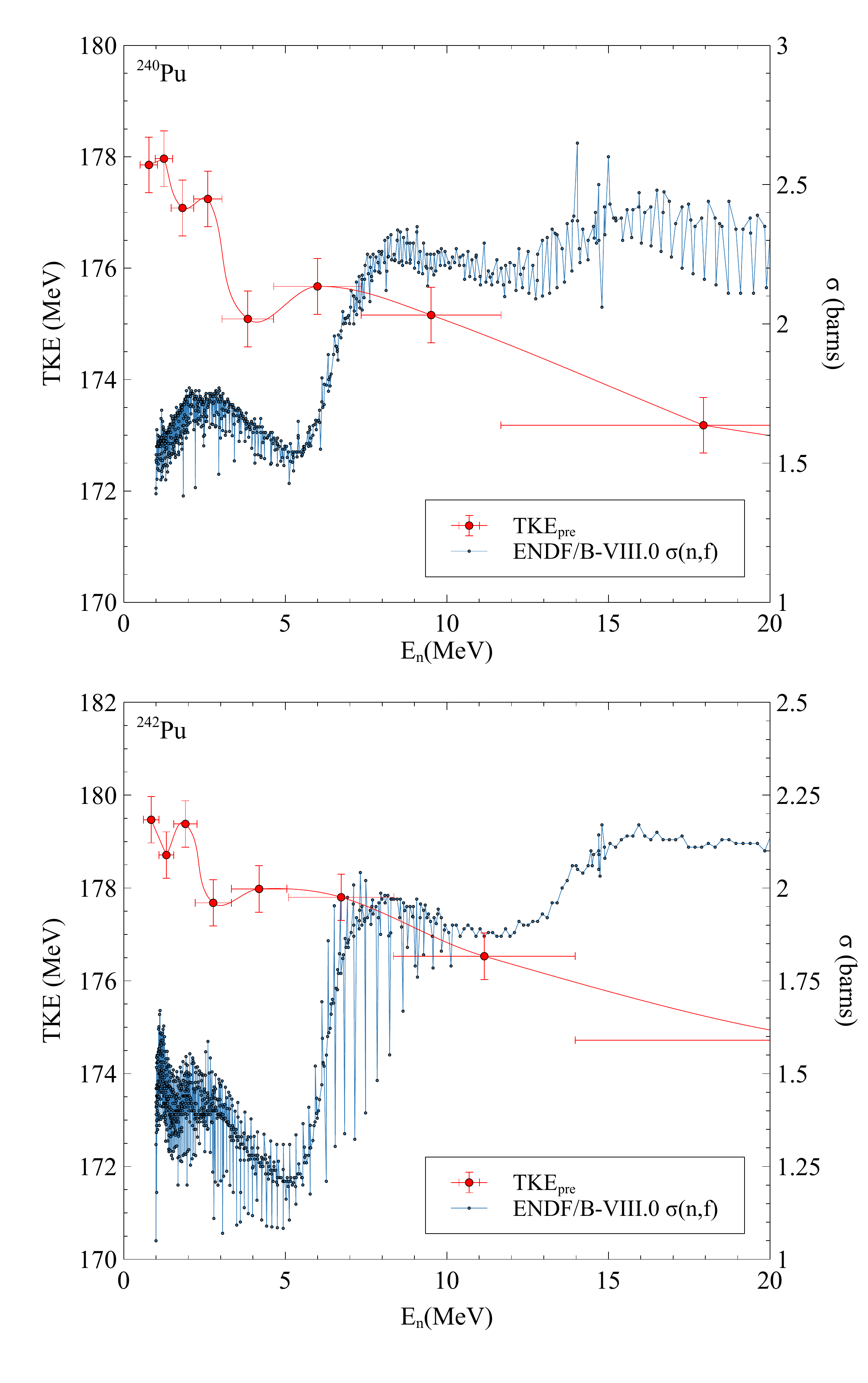}
\caption{The TKE$_{pre}$ plotted as a function of neutron energy for $^{240}$Pu (top) and $^{242}$Pu (bottom) compared to the neutron-induced fission cross section $\sigma$(n,f) from ENDF/B-VIII.0 \cite{ENDF2}.(Color online)}
\label{fig:CrossSection}       % Give a unique label
\end{figure}

Multichance fission in the $^{240,242}$Pu(n,f) systems is possible when the incident neutron energy is greater than 5 MeV \cite{ENDF2}.  In Figure \ref{fig:CrossSection} we show the TKE$_{pre}$ v. E$_n$ relationship compared to the neutron induced fission cross section $\sigma$(n,f) from ENDF/B-VIII.0 \cite{ENDF2}.  In this figure we see an initial rise in TKE$_{pre}$ at the onset of second chance fission.  We attribute the initial rise in TKE$_{pre}$ to the pre-scission neutrons carrying away some energy resulting in less excitation at scission of the compound nucleus. If the energy of the bombarding neutron is increased further, past 5 MeV, but not high enough to overcome the third chance fission threshold, the TKE$_{pre}$ drops. We attribute this behavior to a greater population of energy states for the compound nucleus, the excess energy of which is carried away by prompt gammas and neutrons.  While this phenomenon is apparent in our data, it is damped by the large energy bin widths. 

\subsubsection{Comparison to the Denisov and Sedykh model}
Denisov and Sedykh developed a simple model within the average total kinetic energy $\langle$TKE$\rangle$ framework to highlight the energy dependence of the average kinetic energies of the fission fragments \cite{Denisov1,Denisov2}.  The two key dependencies in the fissioning system upon which the Denisov and Sedykh model is based are:
\begin{enumerate}

\item The width of the fission fragment mass distribution $\sigma^2(E^*)$ is correlated to the excitation energy E$^*$ of the two nascent fragments at the saddle point.
\item The average total kinetic energy $\langle$TKE$\rangle$ depends directly on the width of mass distribution of the fission fragments, $\sigma^2(E^*)$

\end{enumerate}

The following equations explore these two dependencies and describe how we used the model derived by  Denisov and Sedykh to fit our $^{240,242}$ Pu(n,f) measurements.(The complete derivation for the expression for the dependence of the average total kinetic energy $\langle$TKE$\rangle$ of fission fragments on the excitation energy of the compound nucleus can be found in \cite{Denisov1,Denisov2}.)

The excitation energy of the fissioning system at the saddle point,  E$^*$ is assumed to be

\begin{equation}
E^*\approx E^*_{CN} = E_n +B_n
\label{eq:Den1}
\end{equation}

\noindent where E$_{CN}$ is the excitation energy of the compound nucleus, E$_n$ is the energy of the incident neutron and B$_n$ is the binding energy of the neutron.  The value of  B$_n$ for each Pu nuclide was taken from \cite{Wang2017}.

The width of mass distribution $\sigma^2(E^*)$, of fission fragments is calculated by,

\begin{equation}
a_0 = \alpha A + 2^{\frac{1}{3}}\beta A^{\frac{2}{3}}   ,      \kappa = \frac{4 \cdot 2^{\frac{2}{3}}}{9} \beta
\label{eq:Den2}
\end{equation}

\begin{equation}
\sigma^2(E^*) = \frac{2(E^*/a_0)^{1/2}}{C+2 \kappa A^{2/3}E^*/a_0}
\label{eq:Den3}
\end{equation}

\noindent where A is the mass of the fissioning nucleus. The values of the volume and surface level density parameters $\alpha$ = 0.0722396 MeV$^{-1}$ and $\beta$ = 0.195267 MeV$^{-1}$, respectively, were fixed for the calculation and taken from \cite{Capote}.  The value for the stiffness parameter \textit{C}, was obtained by fitting the post-neutron emission TKE data. The values of the stiffness parameter \textit{C} for the fragment asymmetry in the compound nuclei $^{241,243}$Pu were 37.1 and 49.7 MeV respectively.

Finally, the average total kinetic energy \textit{$\langle$TKE$ \rangle$} is given by \cite{Denisov2}

\begin{equation}
\begin{split}
\langle TKE\rangle = k_0 \frac{Z^2}{A^{1/3}}\biggl[1 - \frac{16 \sigma^2(E^*)}{9} \times \\ 
 \left(1 - \frac{exp [ \frac{-1}{4 \sigma^2(E^*)}]}{\sqrt{\pi} \sigma(E^*)   erf [ \frac{1}{2\sigma (E^*)}]}\right)\biggr]
 \end{split} 
\label{eq:Den4}
\end{equation}

\bigskip

\noindent where $k_{0} = \frac{2^{1/3}e^2}{8r_s}$, parameter $r_s$, the distance between the mass center of the fragments at the saddle point, is 1.780 fm \cite{Denisov1} and $\sigma^{2}(E^{*})$ is the fission fragment mass distribution width for a given excitation found by Eq. \ref{eq:Den3}.

\begin{figure}[]
 \includegraphics[trim = {0 0 0 0},clip, width=.9 \columnwidth]{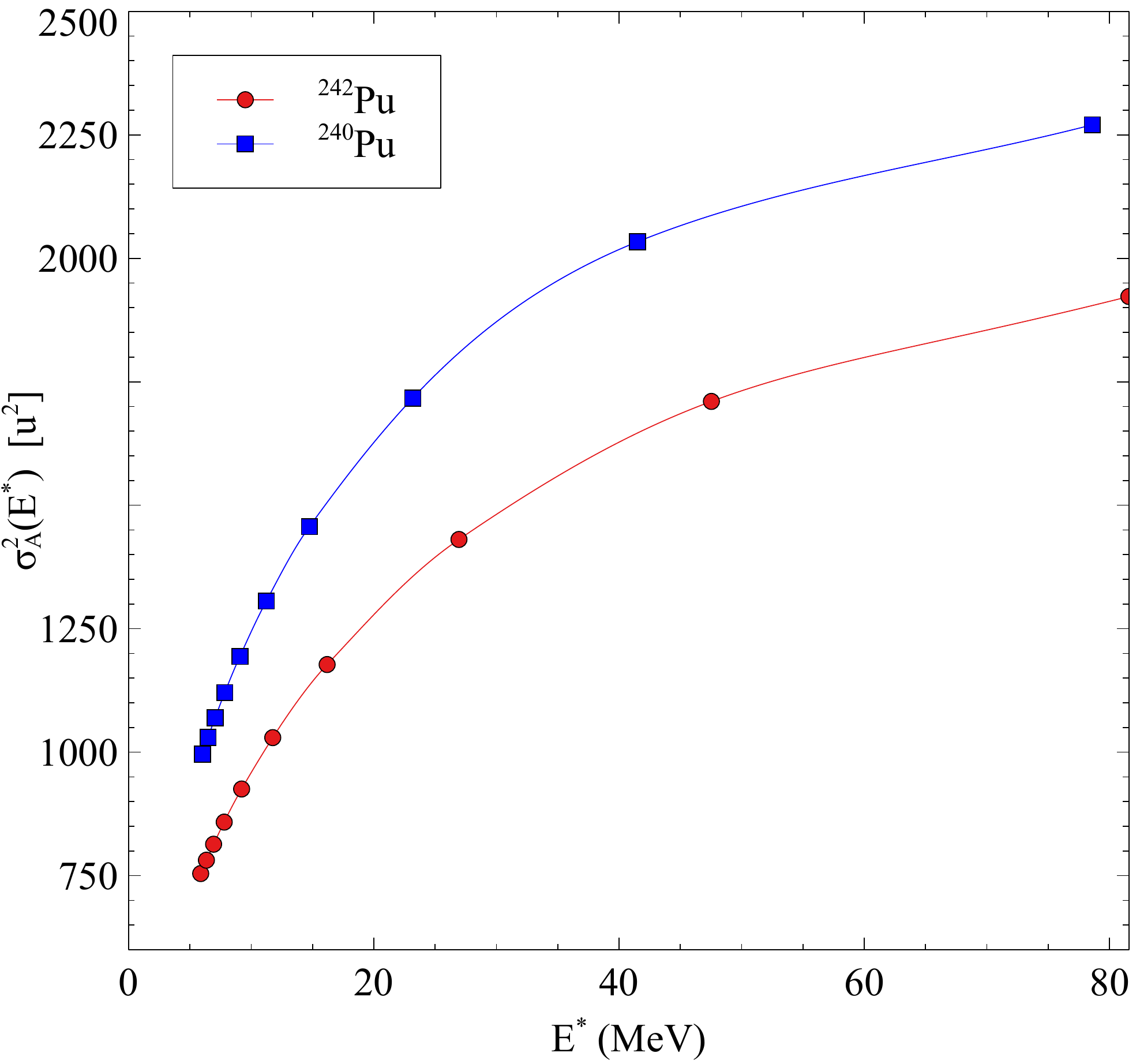}
 \centering
\caption{ The width of the fission fragment mass distribution $ \sigma_A^2(E^*)$ as a function of A for $^{240,242}$Pu(n,f), evaluated using equation \ref{eq:Den3}. \cite{Denisov1,Denisov2}. (Color online)}
\label{fig:MassWidth}       % Give a unique label
\end{figure}

 In Figure \ref{fig:MassWidth} we show fission fragment mass width \textit{$ \sigma_A^2(E^*)$} as a function of A for compound nuclei $^{241,243}$Pu calculated using Eq. \ref{eq:Den3}.  The width of the fission fragment mass yield rises with higher excitation \textit{E$^*$}.  Consequently,  the influence of asymmetric fission events become progressively more significant at the highest excitations \textit{E$^*$}.  We will explore this further in the section on fission channel symmetry.
 
\begin{figure}[h!]
 \includegraphics[trim = {0 0in 0 0in},clip, width=1 \columnwidth]{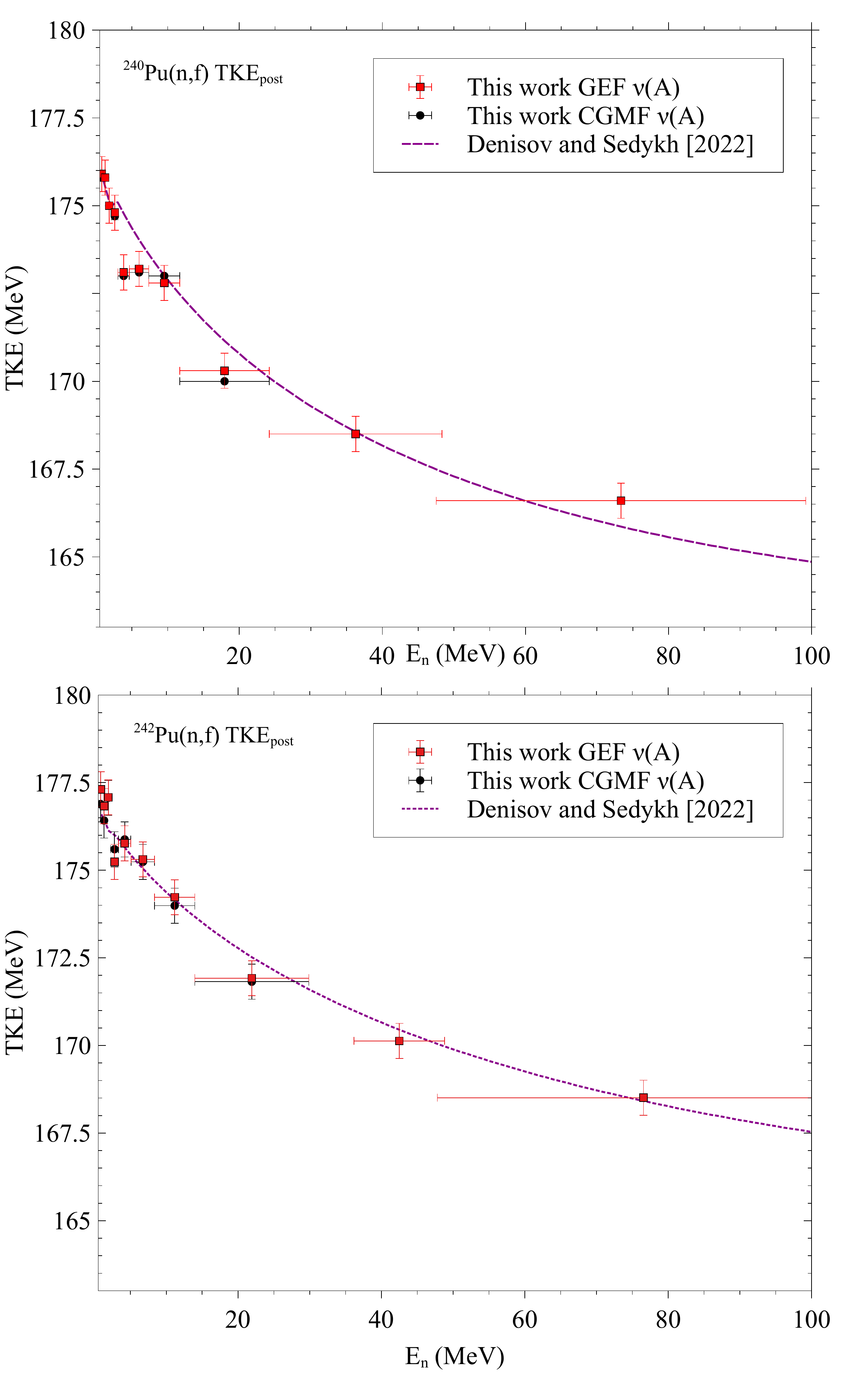}
\caption{Comparison of the dependence of TKE$_{post}$ on incident neutron energy for $^{240}$Pu(n,f) (top) and $^{242}$Pu(n,f) (bottom)calculated using the Denisov and Sedykh model \cite{Denisov2}. The optimized values for stiffness parameter \textit{C} were 37.1 and 49.7 MeV respectively.  (Color online)}
\label{fig:PuTKEDenisov}       % Give a unique label
\end{figure}

In Figure \ref{fig:PuTKEDenisov} we compare the dependence of the average total kinetic energy \textit{$\langle$TKE$\rangle$} on incident neutron energy derived from the Denisov and Sedykh model  \cite{Denisov1,Denisov2} to our experimental data for $^{240,242}$Pu (n,f). The Denisov and Sedykh model well describes our data at energies where E$^* \geq B_n$.  For both the Pu isotopes the calculated average kinetic energy \textit{$\langle$TKE$\rangle$} decreases smoothly with increasing excitation.  Denisov and Sedykh attribute this \textit{$\langle$TKE$\rangle$} decrease to the following two factors; the width of the mass distribution of the fission fragments increases with higher excitation \textit{E$^*$},  and a wider mass distribution indicates asymmetric fission is playing a more dominant role. 

%\begin{figure}[!htb]
%% Use the relevant command to insert your figure file.
%% For example, with the graphicx package use
% \includegraphics[trim = {0 0 0 0},clip, width=.6 \textwidth]{240PuTKEDenisovpost}
%\caption{ Comparison of the dependence of TKE$_{post}$ on incident neutron energy for $^{240}$Pu(n,f) calculated using the Denisov model \cite{Denisov2}. The optimized value for stiffness parameter \textit{C} was 37.1}
%\label{fig:240PuTKEDenisov}       % Give a unique label
%\end{figure}
%
%\begin{figure}[!htb]
%% Use the relevant command to insert your figure file.
%% For example, with the graphicx package use
% \includegraphics[trim = {0 0 0 0},clip, width=.6 \textwidth]{242PuTKEDenisovpost}
%\caption{ Comparison of the dependence of TKE$_{post}$  on incident neutron energy for $^{242}$Pu(n,f) calculated using the Denisov model \cite{Denisov2}. The optimized value for stiffness parameter \textit{C} was 50.5}
%\label{fig:242PuTKEDenisov}       % Give a unique label
%\end{figure}

\subsection{Mass yield distributions}

\begin{figure*}
 \includegraphics[width=1 \textwidth]{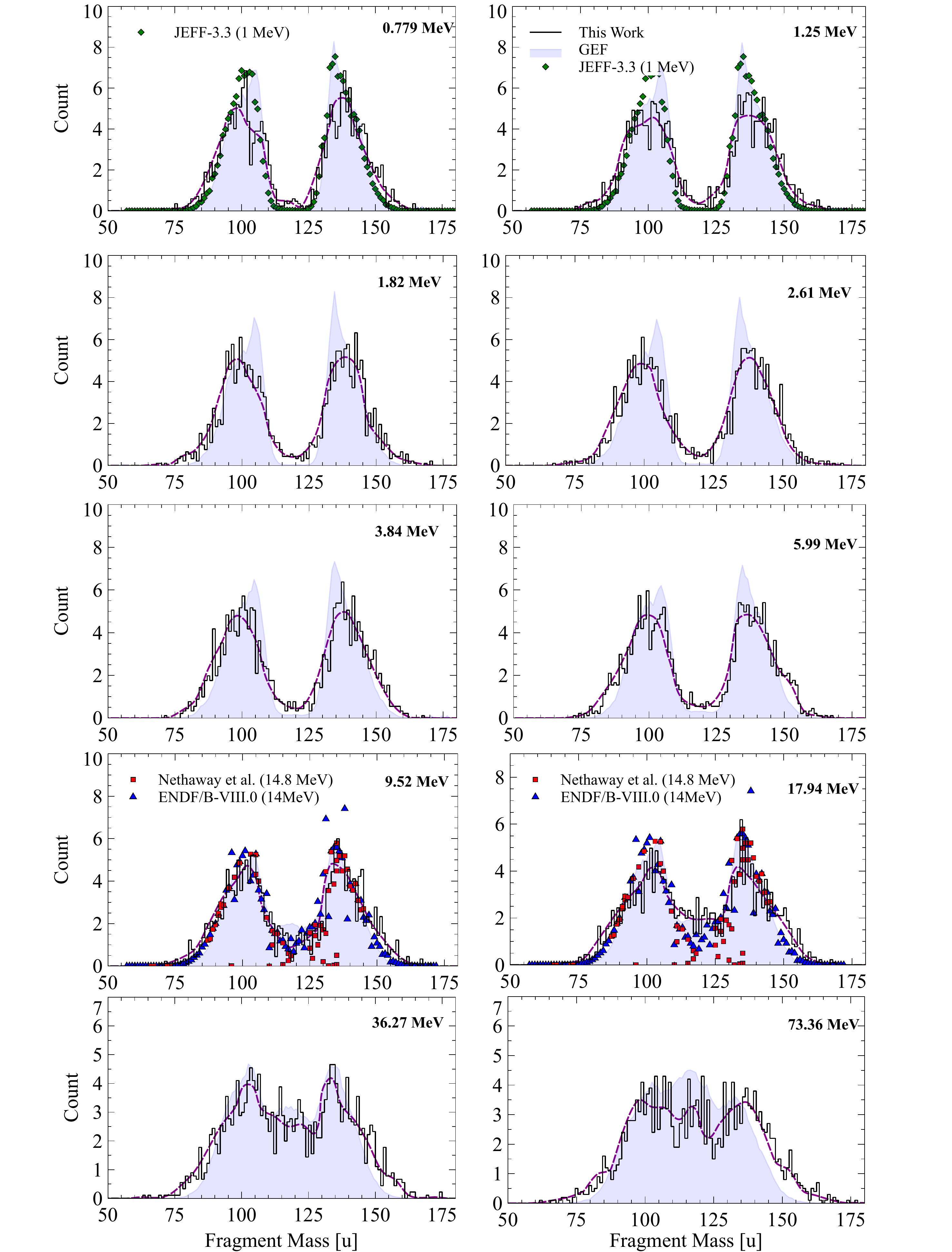}
\caption{Mass yield distribution for the post-neutron evaporation mass yields for $^{240}$Pu(n,f) normalized to 200 $\%$ compared to the predictions of the GEF model \cite{GEF1}.  The smoothed average ($\pm$ 5 u) experimental mass yields are shown by the maroon dashed line.  At select energies we show the evaluated data from \cite{Mills}, \cite{ENDF}, and the experimental data of \cite{Nethaway}.(Color online)}
\label{fig:240MassDistro}       % Give a unique label
\end{figure*} 

\begin{figure*}
\includegraphics[width=1\textwidth]{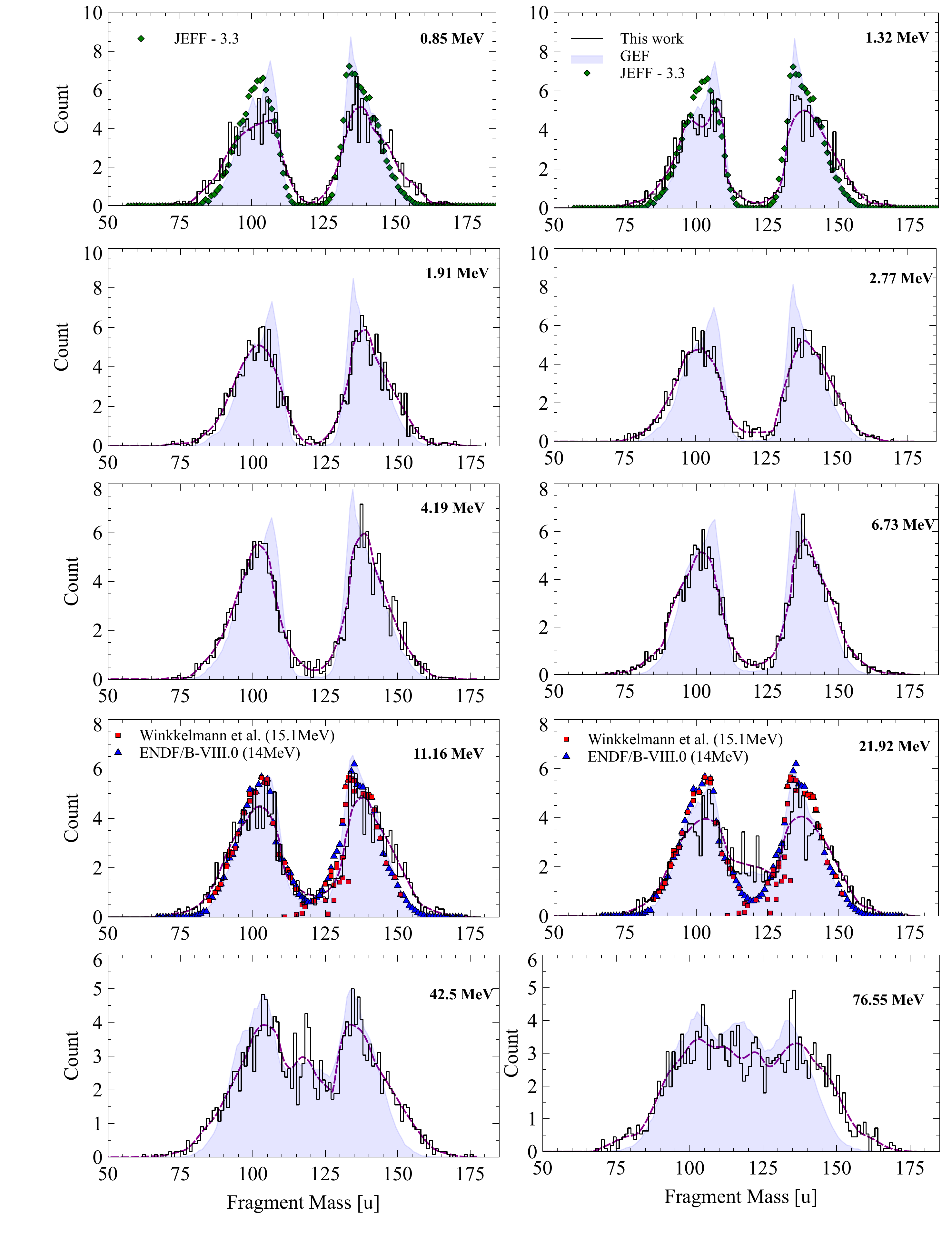}
\caption{Mass yield distribution for the post-neutron evaporation mass yields for $^{242}$Pu(n,f) normalized to 200 $\%$ compared to the predictions of the GEF model \cite{GEF1}. The smoothed average ($\pm$ 5 u) experimental mass yields are shown by the maroon dashed line. At select energies we show the evaluated data from \cite{Mills}, \cite{ENDF}, and the experimental data of \cite{Winkelmann}. (Color online)}
\label{fig:242MassDistro}       % Give a unique label
\end{figure*}

Conservation of momentum and nucleon number dictates that the kinetic energies of the fission fragments are inversely proportional to their masses. Therefore, although not directly measured, mass distributions can be deduced from the kinetic energy of the coincident fragment pairs. 

In Figures \ref{fig:240MassDistro} and \ref{fig:242MassDistro} we plot the post-neutron evaporation mass distributions calculated in this work using the GEF neutron multiplicities (the smoothed average ($\pm$ 5 u) experimental mass yields are shown by the dashed line), as well as the distribution predicted by the GEF model \cite{GEF1}.  We also show the evaluated data from \cite{ENDF} and \cite{Mills} at energies where data is available, as well as the experimental data of \cite{Nethaway} at 14.8 MeV for $^{240}$Pu(n,f) and \cite{Winkelmann} at 15.1 MeV for $^{242}$Pu(n,f). The resolution of our measurement is $\pm$ 5 u.  In both figures it is evident that as the excitation energy of the fissioning system increases, the valley between the asymmetric mass peaks begins to fill in, reflecting the emergence of symmetric fission.  We also see that as the excitation energy is increased, Y$\langle$A$_L\rangle$ and Y$\langle$A$_H\rangle$ decrease due to higher symmetric fission yield. Additionally, the rate of symmetric in growth and its relative proportion to the total number of events are consistent with that predicted by GEF \cite{GEF1}.  That said, while the light fragment mass peak is well described by the GEF model for both Pu systems, our measurements consistently reflect a broader heavy fragment mass peak at the highest energies. 

\begin{figure}
\centering
 \includegraphics[trim = {0 0 0 0},clip,width=.9 \columnwidth]{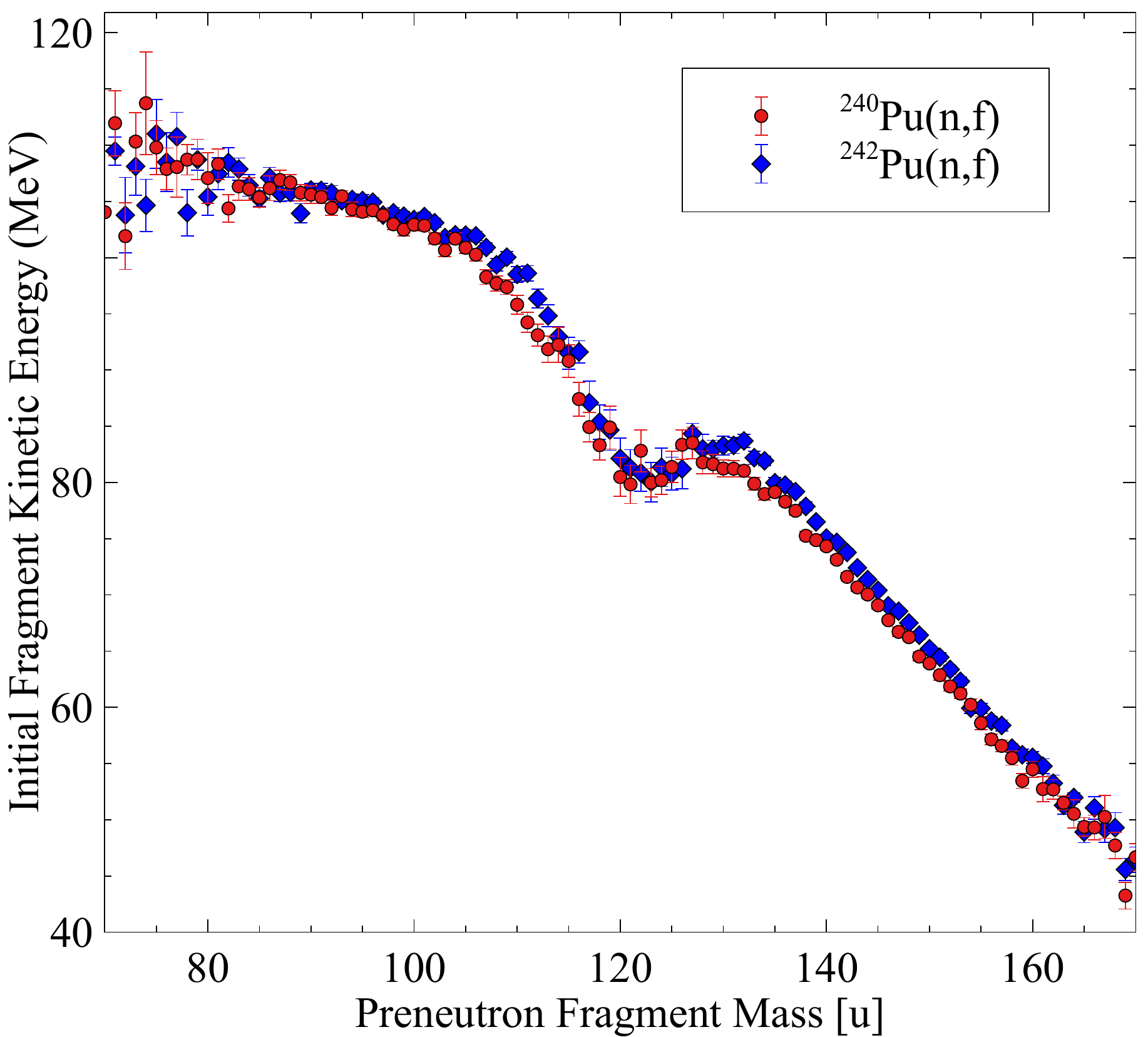}
\caption{The initial fragment kinetic energy (MeV) v.  single fragment mass [u] for $^{240}$ Pu (n,f) and $^{242}$ Pu.  (n,f) for E$_n <$ 5 MeV.(Color online) }
\label{fig:ffTKE3}       % Give a unique label
\end{figure}

Figure \ref{fig:ffTKE3} gives the fragment KE v. single  fragment mass [u] for the two fissioning systems studied. The curves show that the light fragment KE is almost constant while the heavy fragment KE decreases dramatically with increasing mass.  The dip around 120 u is characteristic of symmetric fission.

 \subsection{Fission Channel Symmetry}

\begin{figure}
\centering
\includegraphics[width=.9 \columnwidth]{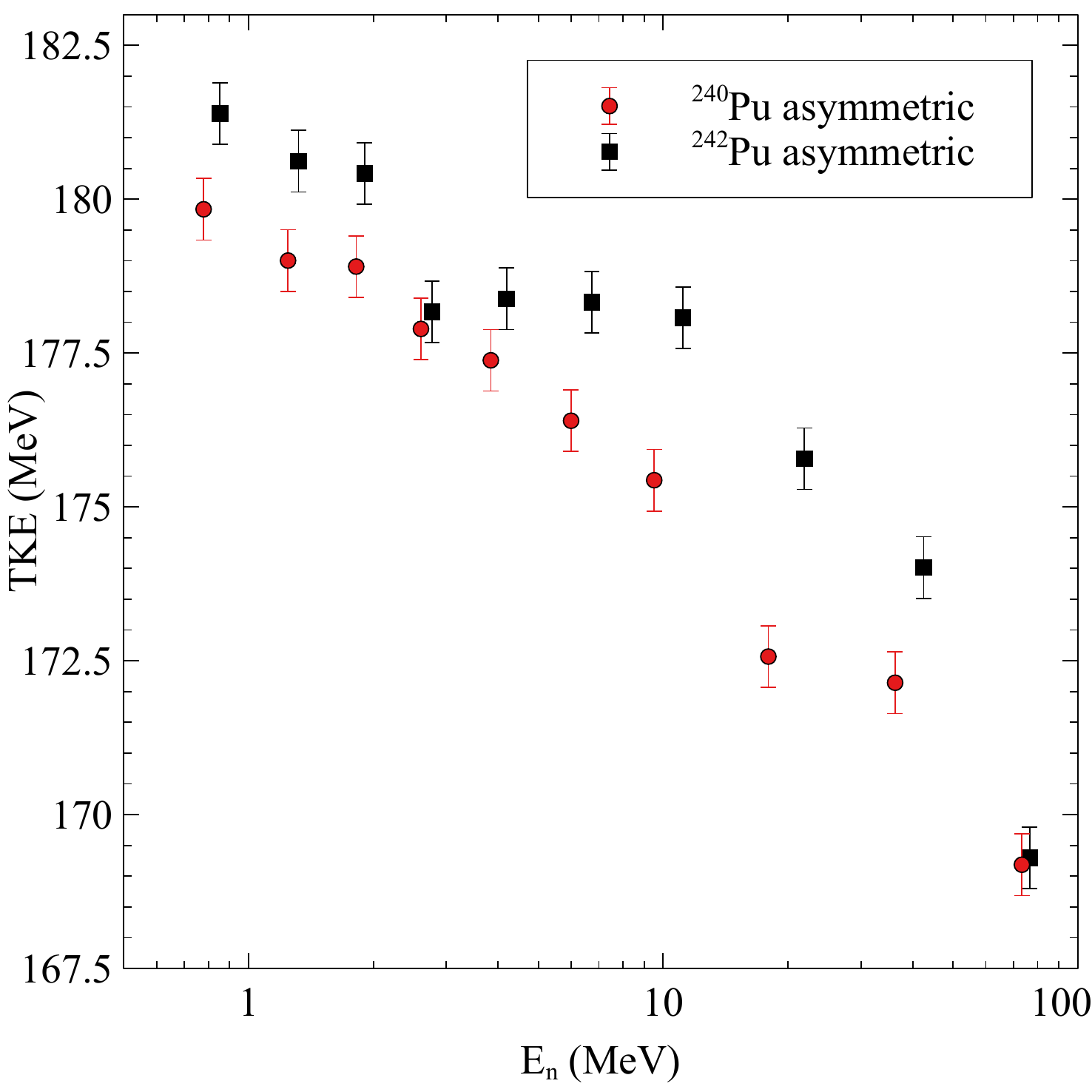}
\caption{Dependence of asymmetric TKE events on neutron energy for the $^{240, 242}$Pu(n,f) reaction.(Color online)}
\label{fig:Asym}       % Give a unique label
\end{figure}

The fission process is dependent on the delicate interplay of microscopic shell effects, and macroscopic nuclear properties.  As E$_n$ increases, it is the dynamics between diminishing shell effects and energy sorting that dictate the fission exit channel.  For actinide fission, asymmetric fission dominates at low energies due to the influence of strong shell effects.However,  as E$_n$ increases the intrinsic energy of the system also rises, giving the fissioning system a wider variety of shapes to embody during its evolution towards fission \cite{Albertsson}. The potential energy landscape at the moment of scission determines the fission exit channel. In Figure \ref{fig:Asym} we plot only the TKE associated with asymmetric fission (TKE$_{asym}$) events for $^{240, 242}$Pu (n,f).  For an event to fall into the asymmetric bin it had to meet the following condition: 

\begin{equation}
 A_{range} = \left(\frac{A_{CN} - v_{(A, E_n)}}{2} + 20 \right) \pm 5 u
 \end{equation}

TKE$_{asym}$ drops off rather steeply, particularly above 5.5 MeV for both $^{240, 242}$Pu (n,f) reactions. The asymmetric fission events in the highest energy bins originate from the multichance fission chain when lighter, less excited CN fission (See Tables \ref{tab:MCF1} and \ref{tab:MCF2}).  The sharp descent in TKE$_{asym}$ has been attributed to the fragments taking on more drastic deformation shapes at higher excitations, particularly for 130 $ < A_H < $ 145 u \cite{Hambsch,Chemey}.  

\subsubsection{Fission fragment distortion and TKE}

\begin{figure}
%% Use the relevant command to insert your figure file.
%% For example, with the graphicx package use
\includegraphics[width=1 \columnwidth]{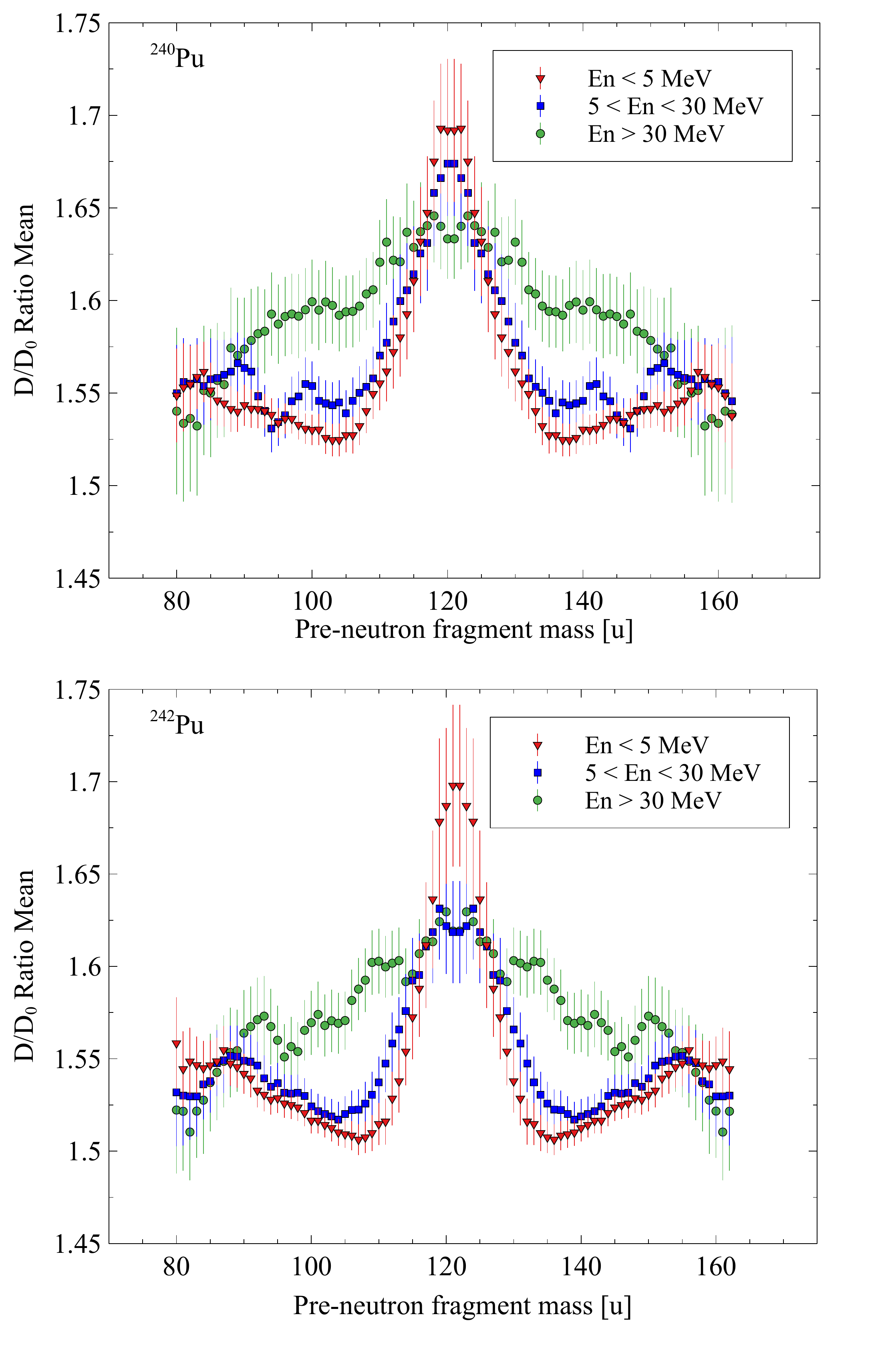}
\caption{Rolling average ratio of the distance between charge centers at scission and that of spherical touching nuclei (D/D$_0$) for $^{240}$Pu (top) and $^{242}$Pu (bottom). (Color online) }
\label{fig:PuD0}       % Give a unique label
\end{figure}

\begin{figure*}
%% Use the relevant command to insert your figure file.
%% For example, with the graphicx package use
\includegraphics[width=1 \textwidth]{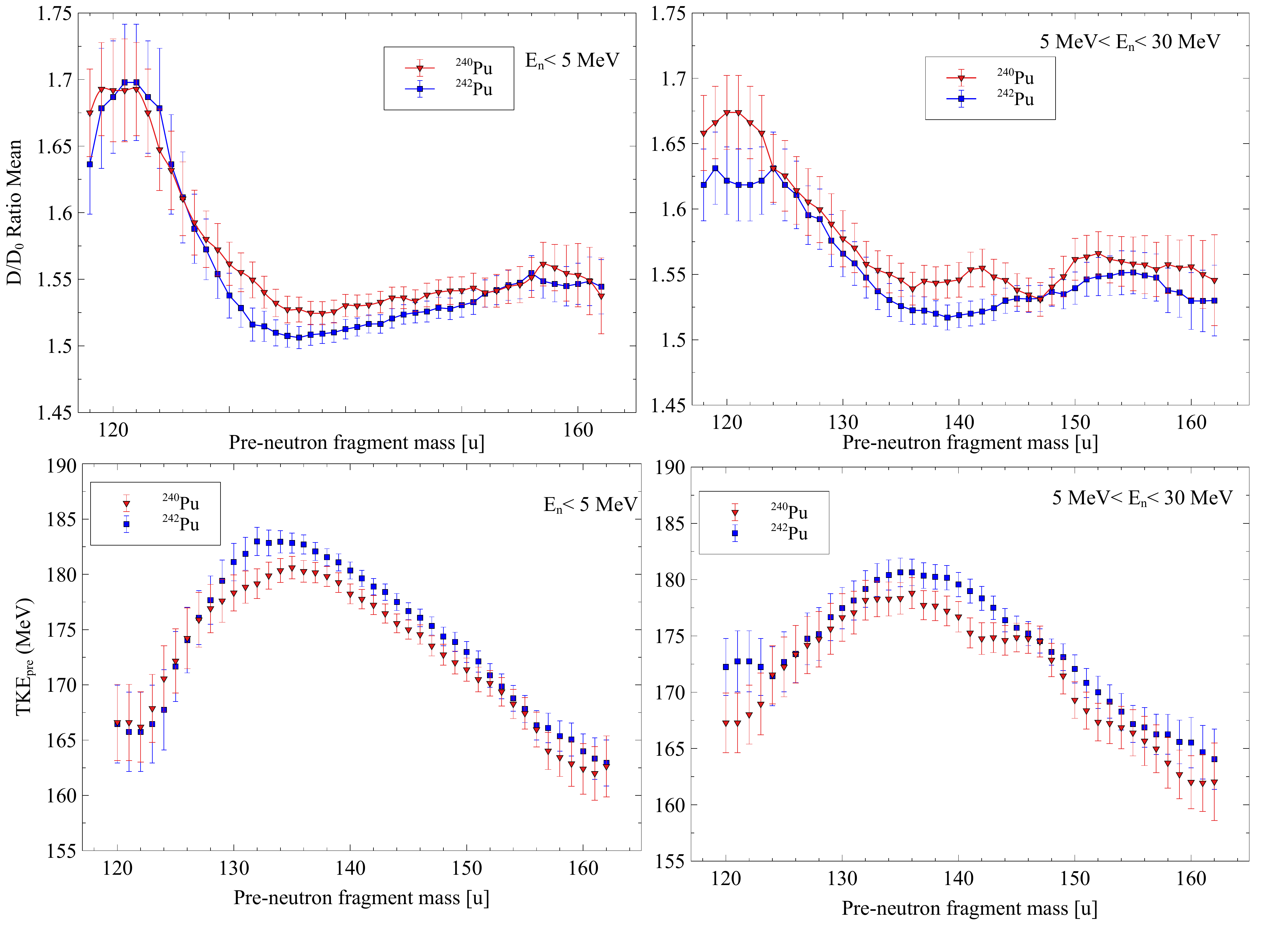}
\caption{(Top) Rolling average ratio of the distance between charge centers at scission and that of spherical touching nuclei (D/D$_0$) for two energy ranges compared to the TKE$_{pre}$ release for $^{240}$Pu and $^{242}$Pu (bottom). (Color online) }
\label{fig:PuD2}       % Give a unique label
\end{figure*}

\begin{figure}
%% Use the relevant command to insert your figure file.
%% For example, with the graphicx package use
\includegraphics[width=1 \columnwidth]{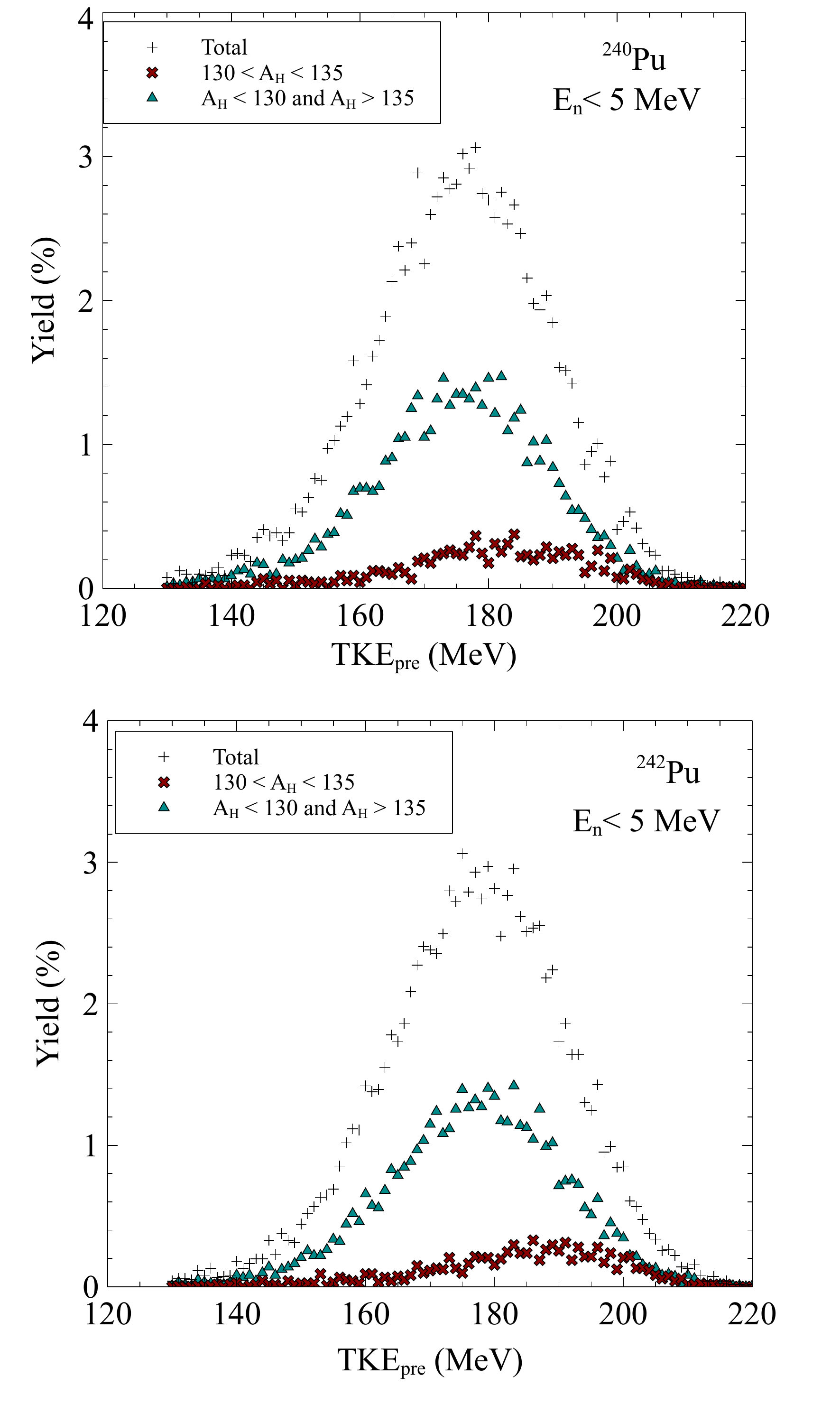}
\caption{TKE$_{pre}$ distributions split into two mass intervals for $^{240}$Pu and $^{242}$Pu. (Color online) }
\label{fig:TKEYield}       % Give a unique label
\end{figure}

\begin{table*}
\caption{Average TKE$_{pre}$ for several mass intervals.}
\centering
\setlength{\tabcolsep}{10pt}
\begin{tabular}{ c ccc}
\hline\hline
Isotope & A$_{H} <$ 130, A$_{H} >$ 135 (MeV)&130 $<$A$_{H} < $ 135 (MeV)& Total (MeV) \\ [0.5ex] % inserts table %heading
\hline\hline
 240	&175.0(2)	&180.5(5)	&174.8(3)\\
242	& 176.4(2)	&184.9(5)	&176.6(3)\\
\noalign{\smallskip}\hline\hline
\end{tabular}
\label{tab:TKEYield}
\end{table*}

Fission is driven by Coulomb repulsion where the magnitude of the energy released is dictated by the distance between the nascent fragments.  We can thus approximate TKE by:

\begin{equation}
TKE(MeV) \approx Z_1Z_2e^2/[D(A_1^{1/3} + A_2^{1/3})]
\label{eq:D0}
\end{equation}

\noindent where e$^2$ is the squared elementary charge (1.44 MeV-fm) and D is the distance between the proto-fragments \cite{Viola2}. Using the relationship described by Eq. \ref{eq:D0} and the Unchanged-Charge-Density (UCD) assumption we can make a reasonable determination of the inter-nuclear distance.  The ratio of the distance between charge centers (D) and the nominal contact distance between touching spherical nuclei (D$_0$) is typically represented as D/D$_0$  where the higher the D/D$_0$  ratio, the greater the fragment deformation \cite{D0}.  In Figure \ref{fig:PuD0} we plot the mean D/D$_0$ ratio as a function of pre-neutron emission fragment mass [u] at various energy ranges.  The fragment mass is plotted as a rolling average of $\pm$ 5 u to correspond with the uncertainty of our measurement. The most compact shapes, i.e. the least deformed fragments,  have D/D$_0$  $\sim$ 1.5 \cite{D0}. For both Pu systems this D/D$_0$ ratio is found in the asymmetric 105 $<$ A$_L$ $<$ 110 u and 130 $<$ A$_H$ $<$ 135 u mass regions.  Furthermore, we see a clear distinction between asymmetric and symmetric fission modes at low E$_n$ with symmetric fragments experiencing a higher degree of deformation.  In the intermediate energy range, the gap between the asymmetric and symmetric modes fills in slightly indicating a marginal increase in asymmetric deformation. At the highest E$_n$, the separation between these fission modes is less distinct with the middle masses having only a slightly higher D/D$_0$ ratio than the mass extremes.  Notably, it is the asymmetric mode that becomes progressively more distorted as E$_n$ increases while there is little variation in the magnitude of symmetric distortion. 

In Figure \ref{fig:PuD2} we compare the relationship between the D/D$_0$ ratio and pre-neutron emission fragment mass for $^{240,242}$Pu  at E$_n < 5$ MeV and 5 $< E_n <$ 30 MeV. Below that we make the comparison to the corresponding TKE$_{pre}$ value.  There is a strong anti-correlation between D/D$_0$ and TKE$_{pre}$. The lowest degree of distortion and highest TKE release is associated with the heavy fragment mass range 130 $ < A_H < $ 135 u for E$_n < 5$ MeV.These fragments are predicted to have a compact spherical shape at scission and therefore experience the most Coulomb repulsion.Contrarily, the highest degree of distortion is associated with the (symmetric) 118 $ < A_H < $ 123 u mass region. Greater deformation increases the fragment intrinsic energy and lowers the available energy to transfer to the fragments at scission, resulting in lower TKE.  In the higher 5 $< E_n <$ 30 MeV energy bin, symmetric fragments experience less deformation and higher TKE than at lower energy, while we see the reverse trend for the asymmetric fragments. This observation is evidence that the sharp decrease in TKE$_{asym}$(shown in Figure \ref{fig:Asym}) is the result of asymmetric fragments experiencing enhanced deformation at higher energies. 

In Figure \ref{fig:TKEYield} the TKE$_{pre}$ distributions are decomposed into two mass regions; 130 $ > A_H $ plus $A_H > $ 135 u, and 130 $ < A_H < $ 135 u.  The average TKE$_{pre}$ values are given in Table \ref{tab:TKEYield}.  For both $^{240}$Pu and $^{242}$Pu in the interval 130-135 u, the average TKE$_{pre}$ is significantly higher, 180.5 and 184.9 MeV respectively, compared to 175.0 and 176.4 MeV in the other interval.  This is because the spherical N = 82 shell will have its maximum influence in the 130 $ < A_H < $ 135 u mass region where stabilization is enhanced by the Z = 50 shell \cite{Wagemans4}.  As the incident neutron energy is increased this stabilizing shell structure is washed out and the yield of 130 $ < A_H < $ 135 u diminishes as symmetric fission becomes more prevalent.

\section{Conclusions}
In this work we investigate the fast neutron induced fission of  $^{240,242}$Pu for which experimental data is extremely scarce.  Thin, highly uniform, vapor deposited $^{240,242}$PuF$_4$ targets were irradiated at LANSCE-WNR to study the dependence of TKE release on incident neutron energies from 1-100 MeV. In addition to TKE$_{pre}$ and TKE$_{post}$ measurements, mass yield distributions were deduced using the 2E method.  We gained additional detail about the $^{240,242}$Pu(n,f) fissioning systems by separating events by symmetry regime to look at the fission exit channel and degree of fragment deformation. Our analysis indicates that the observed TKE decrease with increasing E$_n$ is a consequence of two factors: shell effects fade out at high excitation energies, resulting in the increasing occurrences of symmetric fission, and TKE$_{asym}$ decreases rapidly at high E$_n$ due to more dramatic fragment deformation shapes. 

\section{Acknowledgements}
We thank the Fission TPC Collaboration for their support of the dual glove box system used to manufacture the $^{240, 242}$Pu targets.  We thank S. Kuvin and H.Y. Lee for their assistance in making the measurements and M. Silveira for his assistance in preparing the $^{240,242}$Pu targets. We also thank Amy Lovell for her help running CGMF. This material is based upon work supported in part  by the U.S. Department of Energy, Office of Science, Office of Nuclear Physics under award number DE-FG06-97ER41026 (OSU) and contract number 89233218CNA000001 (LANL).  The nuclide(s) used in this research were supplied by the United States Department of Energy Office of Science by the Isotope Program in the Office of Nuclear Physics.

\bibliography{PuBib.bib}
\bibliographystyle{apsrev4-1}

\end{document}